\documentclass[aps,nofootinbib,longbibliography,prd,onecolumn]{revtex4-1}
\usepackage[babel]{csquotes}
\usepackage{graphicx}
\usepackage{amsmath,amssymb}
\usepackage[colorlinks,citecolor=blue,linkcolor=blue,urlcolor=blue]{hyperref}
\usepackage{mathrsfs}
\usepackage{scalerel}[2016/12/29]
\usepackage{enumerate}
\usepackage{dsfont}
\usepackage{mathtools}
\usepackage{verbatim} 
\usepackage{scrextend} 
\usepackage{float}
\usepackage{subfigure}
\usepackage[utf8]{inputenc} 

\usepackage{xcolor}
\usepackage{xspace}
\usepackage{ifthen}

\newcommand{\mc}[1]%
{\ifthenelse{\equal{\showcomments}{true}}%
{{\color{red}{\small \textbf{MC:} #1}}}{\xspace}}%

\newcommand{\ct}[1]%
{\ifthenelse{\equal{\showcomments}{true}}%
{{\color{cyan}{\small \textbf{CT:} #1}}}{\xspace}}%

\newcommand{\showcomments}{true}

\newcommand{\tar}{\,\kern-0.14em{\rightarrow}\!\bullet}
\newcommand{\sou}{\,\kern-0.14em{\leftarrow}\!\bullet}

\let\oldtableofcontents\tableofcontents
\renewcommand\tableofcontents{

	\hrule width\textwidth 	height 1pt
	\hypersetup{hidelinks}
	\oldtableofcontents
	\vspace{0.5cm}
	\hrule width\textwidth height 1pt
}

\usepackage[top=1.9cm, bottom=2.45cm, left=1.75cm, right=1.75cm]{geometry}

\usepackage{soul}

\def\be{\begin{equation}}
\def\ee{\end{equation}}
\def\bra#1{\mathinner{\langle {#1}|}}
\def\ket#1{\mathinner{|{#1}\rangle}}

\DeclareMathOperator{\e}{e}

\newcommand{\dd}{ \mathrm{d}}
\newcommand\largetimes{\raisebox{-.3ex}{\scaleobj{1.8}{\times}}}

\newcommand{\slc}{SL(2, \mathbb{C})}
\newcommand{\no}{\textsf{n}}
\newcommand{\ed}{\textsf{e}}
\newcommand{\fa}{\textsf{f}}
\newcommand{\ve}{\textsf{v}}
\newcommand{\Tr}{\text{Tr}}

\newcommand{\id}{ \mathds{1}}
\newcommand{\innerp}[2]{\ensuremath{\left\langle {#1} \vert {#2} \right\rangle}}
\newcommand{\transpose}{\intercal}
\newcommand{\CP}{\mathbb{C}\text{P}^1}

\DeclareFontFamily{OMX}{yhex}{}
\DeclareFontShape{OMX}{yhex}{m}{n}{<->yhcmex10}{}
\DeclareSymbolFont{yhlargesymbols}{OMX}{yhex}{m}{n}
\DeclareMathAccent{\arc}{\mathord}{yhlargesymbols}{"F3}

\DeclarePairedDelimiter\floor{\lfloor}{\rfloor}

\DeclarePairedDelimiter\CBr{\{}{\}}

\DeclareRobustCommand{\rchi}{{\mathpalette\irchi\relax}}
\newcommand{\irchi}[2]{\raisebox{\depth}{$#1\chi$}} 

\begin{document}

\title{Geometry Transition in Spinfoams}
\author{Marios Christodoulou\thanks{christod.marios@gmail.com}}
\affiliation{Institute for Quantum Optics and Quantum Information (IQOQI) Vienna,
Austrian Academy of Sciences, Boltzmanngasse 3, A-1090 Vienna, Austria}
\affiliation{Vienna Center for Quantum Science and Technology (VCQ), Faculty of Physics,
University of Vienna, Boltzmanngasse 5, A-1090 Vienna, Austria}

\author{Fabio D'Ambrosio\thanks{fabio.dambrosio@gmx.ch}}
\affiliation{Institute for Theoretical Physics, ETH Zurich, Wolfgang-Pauli-Strasse 27, 8093 Zurich, Switzerland}

\author{Charalampos Theofilis}
\affiliation{Department of Physics, School of Applied Mathematical and Physical 
Sciences, National Technical University of Athens, 9 Iroon Polytechniou Str., 
Zografou Campus GR 157 73, Athens, Greece}
\affiliation{Institute for Quantum Optics and Quantum Information (IQOQI) Vienna,
Austrian Academy of Sciences, Boltzmanngasse 3, A-1090 Vienna, Austria}


\newgeometry{top=3cm, bottom=2.45cm, left=3.5cm, right=3.5cm}

\bigskip

\begin{abstract}
\noindent We show how the fixed-spin asymptotics of the EPRL model can be used to perform the spin-sum for spin foam amplitudes defined on fixed two-complexes without interior faces and contracted with coherent spin-network states peaked on a discrete simplicial geometry with macroscopic areas. We work in the representation given in \cite{Han_Krajewski}. We first rederive the latter in a different way suitable for our purposes. We then extend this representation to 2-complexes with a boundary and derive its relation to the coherent state representation. We give the measure providing the resolution of the identity for Thiemann's state in the twisted geometry parametrization. The above then permit us to put everything together with other results in the literature and show how the spin sum can be performed analytically for the regime of interest here. These results are relevant to analytic investigations regarding the transition of a black hole to a white hole geometry. In particular, this work gives detailed technique that was the basis of estimate for the black to white bounce appeared in  \cite{ChristodoulouCharacteristicTimeScales2018}. These results may also be relevant for applications of spinfoams to investigate the possibility of a `big bounce'.

\clearpage
	\vspace{2cm}
	\tableofcontents
\end{abstract}

\maketitle 
\newgeometry{top=2cm, bottom=2.45cm, left=1.75cm , right=1.75cm}

\section{Introduction}\label{sec:Introduction}

Loop Quantum Gravity (LQG) is a background-independent, non-perturbative theory of Quantum Gravity \cite{thiemann_2007,rovelli_2004, rovelli_vidotto_2014, Ashtekar_2021, Ashtekar:2017yom, Dona:2010hm, Perez:2004hj}. One of the features of the theory is that spacetime itself is a quantum object. A main object of study is the probability amplitudes it defines for a given quantum spacetime configuration. 

In the covariant version of the theory the physics of the geometry transition can be treated a la Feynman in the spirit of a Wheeler–Misner–Hawking sum–over–geometries \cite{RevModPhys.29.497, Misner:1973prb, Hawking:1980gf}. A main outcome in this longstanding direction of research during the previous decade has been what has become known as the EPRL spinfoam transition amplitude of (covariant) Loop Quantum Gravity \cite{Engle_2008, Engle:2012yg, Baratin:2011hp, Dupuis:2011wy, Freidel:2007py, Barrett:1999qw}. The main goal of this paper is to bring the transition amplitude to a form suitable for certain physical applications. In particular, we have in mind the black to white hole transition. This paper also serves companion to the work \cite{ChristodoulouCharacteristicTimeScales2018}. The technique presented here in much more detail was applied in that work to estimate the time it takes for a black hole to quantum transition to a white hole geometry. See also the Chronology note at the end of the paper. This analysis may also be relevant for applications of spinfoams to primordial cosmology.

We show here how the fixed-spin asymptotics of the EPRL model can be used to perform the spin-sum for spin foam amplitudes defined on fixed two-complexes without interior faces, contracted with coherent spin-network states that are peaked on a simplicial classical triangulation on which all the faces have macroscopic areas. The analysis relies on extending techniques previously used to calculate the graviton propagator of LQG to boundary conditions that can be encountered in non-perturbative calculations, such as the black to white transition. We will work in the representation given in \cite{Han_Krajewski}. We will first need to rederive the latter in a different way suitable for our purposes, extend it to 2-complexes that have a boundary, and derive its relation to the coherent state representation. We will also need the measure providing the resolution of the identity for Thiemann's state \cite{thiemann_gauge_2001-3,
thiemann_gauge_2001,thiemann_gauge_2001-2,
thiemann_gauge_2001-1} in the twisted geometry parametrization \cite{freidel_twisted_2010,freidel_twistors_2010} which we derive.

The paper is organized as follows. In Section {\bf II} we briefly review Thiemann's coherent states, derive the integration measure in the twisted geometry parametrization, and set up the approximations relevant to conditioning the transition amplitude with a boundary state peaked on a semi-classical geometry. In Section {\bf III} we rederive the transition amplitude found in \cite{Han_Krajewski} in a way that we can extend it to 2-complexes with boundary and then contract it with coherent states. In Section {\bf IV} we first write the amplitude in the highest weight approximation on a tree-level (no bulk faces) 2-complex. Finally, putting everything together and making use of the semiclassicality conditions set up previously, we show how the spin sum can be approximately but analytically be performed using the stationary phase technique. In Appendix {\bf A} we briefly review $SU(2)$ and $\slc$ representation theory in the notation and conventions used here. In Appendix {\bf B} we explain why in this setting the Jacobi theta function appearing in the final result can simply be replaced by unit.

\section{Semiclassical Boundary States} \label{ssec:SubSection_BoundaryStates}
In this section we briefly review the boundary states used in subsequent sections. In \ref{Section_SemiclassicalResOfId} we derive the measure providing the resolution of identity in the twisted geometry parametrisation which will be needed in what follows.\footnote{This result is also crucial in order to take into account the issue raised by R.~Oeckl in \cite{Oeckl:2018nwg}, where it is suggested that the measure must be considered in the definitions of the observables studied in \cite{christodoulou_realistic_2016,
ChristodoulouCharacteristicTimeScales2018} for a black hole to white hole transition.} In \ref{Area_Limit} we review the large areas limit of the states, and discuss semiclassicality conditions, the validity of which will be the central assumption for performing the spin--sum in Section \ref{sec:SpinSum}.

\subsection{Resolution of Identity in the Twisted Geometry Parametrization for the Heat Kernel States}
\label{Section_SemiclassicalResOfId}
 The boundary states considered throughout this paper are Thiemann's heat kernel states \cite{thiemann_gauge_2001-3,
thiemann_gauge_2001,thiemann_gauge_2001-2,
thiemann_gauge_2001-1}, in the twisted-geometry parametrization \cite{freidel_twisted_2010,freidel_twistors_2010}.  When parametrized in this manner, the states are also known as coherent spin-networks \cite{bianchi_coherent_2010} or extrinsic coherent states \cite{Rovelli_New_Book}. They are elements of the truncated boundary Hilbert space $\mathcal{H}_\Gamma=L^2\left[SU(2)^L / SU(2)^N\right]$ and are labelled by data $H_\ell$ drawn from the discrete phase space $P_\Gamma = \largetimes_\ell T^{*}SU(2)_\ell \simeq \largetimes_\ell \left(\mathbb{R}^+_\ell\times S^1_\ell \times S^2_\ell \times S^2_\ell \right)$ of twisted geometries. Here,  $L$ denotes the number of links $\ell$ and $N$ the number of nodes $\no$ of the boundary graph $\Gamma$. Coherent spin-networks are defined as the $L$-parameter family of states
\begin{equation}\label{eq:ThiemannState}
	\Psi_{\Gamma, H_{\ell}}^{t_\ell}(h_\ell) := \int_{SU(2)^N}\left(\prod_{\no}\,\dd h_{\no(\ell)}\right)\, \prod_{\ell} K_{\ell}^{t_\ell} (h_\ell, h_{t(\ell)}\, H_{\ell}\, h_{s(\ell)}^{-1}),
\end{equation}
where $s(\ell)$ and $t(\ell)$ denote source and target node of the link $\ell$, $t_\ell>0$ are the $L$ semiclassicality parameters and $K^t(h, H)$ is the $SU(2)$ heat kernel with a complexified $SU(2)$ element as second argument. Since $SU(2)^{\mathbb{C}}\simeq SL(2, \mathbb{C})$, $H$ is taken to be an element of $SL(2, \mathbb{C})$\footnote{$SL(2,\mathbb{C})$ is isomorphic to $SU(2) \times su(2) \simeq T^*SU(2)$ which corresponds to the (linkwise, not gauge invariant) classical phase space associated to the Hilbert space on a graph.}. The Wigner D-matrices of the $SU(2)$ heat kernel in \eqref{eq:ThiemannState} are defined by analytical extension to the group $SL(2,\mathbb{C})$\footnote{The explicit defining expression for the analytically extended matrix elements $D^j_{m n}$ can be found in \cite{RuhlBook} and \cite{Barrett_2009}. In fact, this provides an analytic extension to the entire $GL(2,\mathbb{C})$.}. Concretely, $K^{t}(h, H)$ is given in the spin-representation by
\begin{equation}
	K^{t}(h, H) = \sum_{j} d_j \e^{-j(j+1) t} \text{Tr}\left[D^{(j)}(h H^{-1})\right].
\end{equation}
The $L$-parameter family of states $\Psi_{\Gamma, H_\ell}^{t_\ell}(h_\ell)$ is an overcomplete basis of the Hilbert space $\mathcal{H}_\Gamma$. The identity operator $\id_\Gamma$ on $\mathcal{H}_\Gamma$ is given in the holonomy representation by the delta distribution $\delta_\Gamma$ on $SU(2)^L / SU(2)^N$, and we have
\begin{equation}
	\delta_\Gamma(h_l, h'_l) = \int_{SL(2, \mathbb{C})^L}\left(\prod_l\Omega_{2 t_{\ell}}(H_l)\,\dd H_l\right) \Psi^{t_\ell}_{\Gamma, H_l}(h_l)\,\overline{\Psi^{t_\ell}_{\Gamma, H_l}(h'_l)}.
\end{equation}
The twisted geometry parametrization relies on the Cartan decomposition of $H_\ell^{-1}\in \slc$, i.e.
\begin{equation}\label{eq:HParam}
	H_\ell^{-1} = n_{s(\ell)}\,\e^{(\eta_\ell + i \gamma \zeta_\ell)\frac{\sigma_3}{2}}\, n_{t(\ell)}^{-1}.
\end{equation}
The data $H_\ell$ are replaced by the data $(\eta_\ell, \zeta_\ell, \vec{n}_{s(\ell)}, \vec{n}_{t(\ell)})$, which at the classical level have the following geometrical interpretation: the data $\eta_\ell\in \mathbb{R}^+$ is related to the area dual to the link $\ell$, $\zeta_\ell\in [0, 4\pi)$ encodes the distributional extrinsic curvature \cite{Rovelligeometryloopquantum2010}, $\gamma$ is the Barbero-Immirzi parameter and $\vec{n}_{s(\ell)}, \vec{n}_{t(\ell)}$ are two unit vectors normal to the face dual to the link. Substituting \eqref{eq:HParam} into \eqref{eq:ThiemannState} allows the construction of coherent states peaked on a prescribed boundary discrete geometry. 

The measure $\Omega_{2t}(H)$ which provides the resolution of identity in terms of Thiemann's coherent states is formally given as the heat kernel on the quotient space $\slc/SU(2)$, i.e.
\begin{equation}
	\Omega_{2t}(H) := \int_{SU(2)}F_{2t}(H g)\, \dd g, 
\end{equation}
where $F_{2t}$ is the heat kernel on $\slc$. The integration measure in the polar decomposition of $\slc$, 
\begin{equation} \label{eq:polarDecomp}
	H =  h\, \e^{\vec{p}\cdot\frac{\vec{\sigma}}{2}}.
\end{equation}
where $h \in SU(2)$ and $\vec{p}$ is a vector in $\mathbb{R}^3$, is given by \cite{thiemann_gauge_2001-2, bianchi_coherent_2010}
\begin{equation} \label{eq:polarMeasure}
	\Omega_{2t}(h\, \e^{\vec{p}\cdot\frac{\vec{\sigma}}{2}}) = \frac{\e^{\frac{t}{2}}}{(2 \pi t)^{\frac{3}{2}}}\frac{\vert \vec{p} \vert}{\sinh \vert \vec{p} \vert}\e^{-\frac{\vert p \vert^2}{2 t}}\quad\quad\quad \dd H = \frac{\sinh^2\vert\vec{p}\vert}{\vert \vec{p}\vert^2}\,\dd h\,\dd^3\vec{p}
\end{equation}

With these preliminaries, in the remaining of this section we derive the measure in the twisted geometry parametrization. First, note that the $SU(2)$ element $h$ does not appear on the right hand side of $\Omega_{2t}$ in \eqref{eq:polarMeasure}, because it is by definition an $SU(2)$ invariant function. Similarly, for the twisted geometry parametrization \eqref{eq:HParam} we have
\begin{equation}
	\Omega_{2t_\ell}\left(n_{s(\ell)}\,\e^{\eta_\ell\frac{\sigma_3}{2}}\,\e^{i \gamma \zeta_\ell\frac{\sigma_3}{2}}\, n_{t(\ell)}^{-1}\right) = \Omega_{2t_\ell}\left(\e^{\eta_\ell\frac{\sigma_3}{2}}\right) = \frac{\e^{-\frac{t_\ell}{2}}}{(2 \pi t_\ell)^{\frac{3}{2}}}\frac{\eta_\ell}{\sinh \eta_\ell}\,\e^{-\frac{\eta^2_\ell}{2 t_\ell}}
\end{equation}
where again the $SU(2)$ elements $n_{s(\ell)}$ and $n_{t(\ell)}$ drop out from the right hand side because of the $SU(2)$ invariance of $\Omega_{2t}$. The measure $\dd H_\ell$ in the Cartan decomposition reads\footnote{See, for instance, \cite{RuhlBook}.}
\begin{equation}\label{eq:CartanMeasure}
	\dd H_\ell = \frac{\sinh^2 \eta_\ell}{4\pi}\,\dd\eta_\ell\,\dd u_\ell\,\dd v_\ell,
\end{equation}
where $\dd u_\ell$ and $\dd v_\ell$ are $SU(2)$ Haar measures. We seem to have achieved our goal, but, the resolution of identity in the twisted geometry parametrization does not immediately follow from the above expressions because of the following subtlety.  The polar decomposition of  \eqref{eq:polarDecomp} for $H_\ell$ is unique 
and is a parametrization by six real parameters. The twisted geometry parametrization \eqref{eq:HParam} for $H_\ell$ is not unique. There is a $U(1)$ gauge choice to be made, since there are seven real parameters to be integrated over in \eqref{eq:CartanMeasure}. We proceed by ansatz, choosing to drop the $\zeta_\ell$ integration in $\dd u_\ell$ such that the measure becomes proportional to the standard measure on the two-sphere $\mathcal S^2$. The measure $\dd v_\ell$ remains the standard $SU(2)$ Haar measure. Concretely:
\begin{align}
	\dd u_\ell &:= \mathcal N \sin\theta_{s(\ell)}\,\dd \phi_{s(\ell)}\,\dd \theta_{s(\ell)} = \mathcal N\,\dd^2 \vec{n}_{s(\ell)}\,\notag\\
	\dd v_\ell &:= \frac{1}{(4\pi)^2}\sin\theta_{t(\ell)}\,\dd \phi_{t(\ell)}\,\dd \theta_{t(\ell)} \,\dd \zeta_{\ell} = \frac{1}{(4\pi)^2}\dd^2 \vec{n}_{t(\ell)} \,\dd\zeta_{\ell}.
\end{align}
The full ansatz for the resolution of identity measure then reads
\begin{align}\label{eq:MeasureAnsatz}
	\Omega_{2t_\ell}(H_\ell)\,\dd H_\ell = \frac{\mathcal{N}}{(4\pi)^2}\,\Omega_{2t_\ell}\left(\e^{\eta_\ell\frac{\sigma_3}{2}}\right) \,\sinh^2\eta_\ell\,\dd \eta_\ell\,\dd \zeta_\ell\,\dd^2 \vec{n}_{s(\ell)}\,\dd^2\vec{n}_{t(\ell)}.
\end{align} 
We fix the normalization $\mathcal{N}$ by requiring that the ``volume" of this measure over $\slc$ is the same in the polar decomposition and the Cartan decomposition
\begin{equation}\label{eq:MeasureAnsatz}
	\int_{\slc}\mathcal{N}\,\Omega_{2t_\ell}\left(\e^{\eta_\ell\frac{\sigma_3}{2}}\right)\,\sinh^2\eta_\ell\,\dd \eta_\ell\,\dd \zeta_\ell\,\dd^2 \vec{n}_{s(\ell)}\,\dd^2\vec{n}_{t(\ell)} = \int_{\slc}\Omega_{2t_\ell}(\e^{\vec{p}\cdot\frac{\vec{\sigma}}{2}})\,\frac{\sinh^2\vert\vec{p}\vert}{\vert \vec{p}\vert^2}\, \dd h\, \dd^3\vec{p}.
\end{equation}
On the left hand side the two integrations over $\mathcal{S}^2$ and the integration over $\zeta_\ell$ give an overall factor of $(4\pi)^3$ and on the right hand side the $SU(2)$ integration gives unit. There remain the integrations over $\eta_\ell$ and $\vec{p}$. From \eqref{eq:polarMeasure}, the integrand on the right hand side only depends on the norm $p\equiv\vert \vec{p}\vert$. We can therefore change to polar coordinates for the vector $\vec{p}$ which gives another factor of $4 \pi$ from the angular integration. The remaining integrals over $\eta_\ell$ and $p$ are of the same form and they are both non zero since the integrand is positive definite. Hence, collecting all terms and solving for $\mathcal{N}$ we find $\mathcal N = 1$.

We now proceed to show by direct calculation (that is, that the above ansatz indeed works out) that the measure \eqref{eq:MeasureAnsatz} indeed gives the resolution of identity for the states \eqref{eq:ThiemannState} in the twisted geometry parametrization. To simplify the notation and render the computations more readable, we drop the gauge-averaging integrations over $SU(2)$ and only consider a single link. The full proof proceeds similarly. We wish to prove that
\begin{equation}\label{eq:ResOfId}
\delta(hh'^\dagger) = \int_{\mathbb{R}^+}\dd\eta\,\nu_t(\eta)\int_{0}^{4\pi}\dd\zeta\int_{\mathcal{S}^2}\dd^2 \vec{n}_s\int_{\mathcal{S}^2}\dd^2 \vec{n}_t\, \Psi_{H}^t(h)\,\overline{\Psi_H^t(h')},
\end{equation}
where 
\begin{align}
	\delta(hh'^\dagger) = \sum_j d_j \Tr_j[hh'^\dagger] = \sum_j d_j \sum_{|m|\leq j}\sum_{|n| \leq j}D^j_{mn}(h)D^j_{nm}(h'^\dagger)
\end{align}
is the Dirac distribution on $SU(2)$ and $\nu_t(\eta)$ is given by
\begin{align}
	\nu_t(\eta) := \frac{\e^{-\frac{t}{2}}}{(4\pi)^2(2\pi t)^{3/2}} \eta \sinh \eta\,\e^{-\frac{\eta^2}{2t}}.
\end{align}
The states \eqref{eq:ThiemannState} in the twisted geometry parametrization \eqref{eq:HParam} are explicitly given by 
\begin{align}
	\Psi_{\Gamma, H}^t(h) &= \sum_j d_j \,\e^{-j(j+1) t} \sum_{m, n, k} D^j_{mn}(h) D_{nk}^j(n_t) D_{km}^j(\e^{i \gamma\zeta \frac{\sigma_3}{2}}n_s^\dagger)\,\e^{-\eta k}     \notag\\
	\overline{\Psi_{\Gamma, H}^t(h')} &= \sum_{j'} d_{j'}\,\e^{-j'(j'+1) t}\sum_{m', n', k'} D_{n' m'}^{j'}(h'^\dagger)\,D_{k' n'}^{j'}(n_t^\dagger) D_{m' k'}^{j'}(n_s \e^{-i \gamma \zeta \frac{\sigma_3}{2}})\, \e^{-\eta k'}.
\end{align}
By noticing that $n_s = \e^{-i\phi_s\frac{\sigma_3}{2}}\e^{-i\theta_s\frac{\sigma_3}{2}}$ lives in a subspace of $SU(2)$ we can introduce the auxiliary variable $g:=n_s \e^{-i\gamma \zeta \frac{\sigma_3}{2}}$, which is a genuine $SU(2)$ element. This allows us to perform the $n_s$ and the $\zeta$ integration simultaneously by virtue of the Peter-Weyl theorem
\begin{align}
	A &:= \int_{\mathcal{S}^2}\dd^2\vec{n}_s \int_0^{4\pi}\dd \zeta\, \Psi_{H}^t(h)\,\overline{\Psi_H^t(h')} = (4\pi)^2\int_{SU(2)}\dd g \, \Psi_{H}^t(h)\,\overline{\Psi_H^t(h')}\notag\\
	&=(4\pi)^2 \delta^{j j'} \delta_{m m'} \delta_{k k'} \sum_j d_j \e^{-2j(j+1) t} \sum_{m, n, k, n'} D_{mn}^j(h) D_{nk}^j(n_t) D_{n' m}^j(h'^\dagger) D_{k n'}^j(n_t^\dagger) \e^{-2\eta k}.
\end{align}
To perform the next integration we notice that $n_t$ is also parametrized as $n_t = \e^{- i \phi_t\frac{\sigma_3}{2}}\,\e^{-i\theta_t\frac{\sigma_2}{2}}$ and that therefore we have
\begin{align}
	D_{nk}^j(n_t) = \e^{-i\phi_t n} d_{nk}^j(\theta_t), \quad\quad D_{k n'}^j(n_t^\dagger) = \e^{i\phi_t n'}\, d_{k n'}^j(-\theta_t),
\end{align}
which implies that the $n_t$ integration is zero unless $n=n'$. This allows us to do the following step:
\begin{align}
	\int_{\mathcal{S}^2}\dd^2 \vec{n}_t\, D_{nk}^j(n_t) D_{kn'}^j(n_t^\dagger) &=  \delta_{n n'} \int_{\mathcal{S}^2}\dd^2 \vec{n}_t\, D^j_{nk}(n_t) D^j_{kn}(n_t^\dagger)\,\frac{1}{4\pi}\int_0^{4\pi} \dd\xi\, \e^{i\xi k}\,\e^{-i\xi k}\notag\\
	&= (4\pi) \delta_{n n'} \int_{SU(2)} \dd g' D_{nk}^j(g') D_{kn}^j(g'^\dagger) = \frac{4\pi \delta_{n n'}}{d_j}.
\end{align}
Above, we inserted the identity in the first equality and defined the auxiliary variable $g' := n_t\,\e^{-i\xi \frac{\sigma_3}{2}}$, which allowed us to use again the Peter-Weyl theorem. Hence we find:
\begin{align}
	B:=\int_{\mathcal{S}^2}\dd^2\vec{n}_t \,A = (4\pi)^3 \sum_j \e^{-2j(j+1)t}\sum_{m, n} D^j_{mn}(h)D_{nm}(h'^\dagger)\sum_k\e^{-2\eta k}.
\end{align}
The last sum is easily performed by recognizing that it can be split into two geometric sums and yields
\begin{align}
	\sum_{\vert k\vert\leq j} \e^{-2\eta k} = \frac{\sinh\left((2j+1)\eta\right)}{\sinh\eta},
\end{align}
which holds for both, integer and half-integer values of $j$. What is left is the integral over $\eta$ which gives
\begin{align}
	(4\pi)^3\int_{\mathbb R^+}\dd \eta\,\nu_t(\eta)\frac{\sinh\left((2j+1)\eta\right)}{\sinh\eta} = d_j\,\e^{2j(j+1)t}.
\end{align}
Putting everything together we obtain
\begin{align}
	\int_{\mathbb{R}^+}\dd\eta \,\nu_t(\eta) B = \sum_j d_j \sum_{m, n}D_{mn}^j(h) D_{nm}^j(h'^\dagger) = \delta(h h'^\dagger).
\end{align}
This completes the proof. The above steps extend straightforwardly to gauge-invariant states \eqref{eq:ThiemannState} on a general graph $\Gamma$, and one can then also prove the identity
\begin{align}
	\delta_\Gamma(h, h'^\dagger) = \int_{\slc^L} \left(\prod_\ell\nu_{t_{\ell}}(\eta_\ell)\dd\eta_\ell\,\dd\zeta_\ell\,\dd\vec{n}_{s(\ell)}\,\dd\vec{n}_{t(\ell)}\right) \Psi_{\Gamma, H_\ell}^{t_\ell}(h)\overline{\Psi_{\Gamma, H_\ell}^{t_\ell}(h')}
\end{align}
with the Dirac distribution $\delta_\Gamma$ on $SU(2)^L / SU(2)^N$ explicitly given by
\begin{align}
	\delta_\Gamma(h, h'^\dagger) = \int_{SU(2)^N}\left(\prod_\no\dd h_{\no(\ell)}\right) \int_{SU(2)^N}\left(\prod_\no\dd \tilde{h}_{\no(\ell)}\right)\prod_\ell \delta\left(h_{t(\ell)}^\dagger h h_{s(\ell)}\,\left(\tilde{h}_{t(\ell)}^\dagger h' \tilde{h}_{s(\ell)}\right)^\dagger\right).
\end{align}

\bigskip
In summary, the integration measure giving the resolution of the identity for Thiemann's coherent states in the twisted geometry parametrization reads
\begin{align}
	&\Omega_{2t_\ell}(\e^{\eta_\ell\frac{\sigma_3}{2}})\,\dd H_\ell = \frac{\e^{-\frac{t_\ell}{2}}}{(4\pi)^2(2\pi t_\ell)^{3/2}} \eta_\ell \sinh \eta_\ell\,\e^{-\frac{\eta_\ell^2}{2t_\ell}}\,\dd\eta_\ell\,\dd\zeta_\ell\,\dd^2 \vec{n}_{s(\ell)}\,\dd^2 \vec{n}_{t(\ell)}\notag\\
	&\quad\quad\quad\quad\quad\eta_\ell\in \mathbb{R}^+,\quad \zeta_\ell\in [0, 4\pi), \quad\vec{n}_{s(\ell)}\in\mathcal{S}^2, \quad\vec{n}_{t(\ell)}\in\mathcal{S}^2
\end{align}

\subsection{The large Areas Limit}\label{Area_Limit}

In this subsection we briefly review the large area (large $\eta$) limit of the Thiemann's states in the twisted geometry parametrization and the interpretation of the data $(\eta_\ell, \zeta_\ell, \vec{n}_{s(\ell)}, \vec{n}_{t(\ell)})$, as appeared in \cite{bianchi_coherent_2010}. This discussion also provides the kinematical setup and assumptions under which we perform the spin--sum in Section \ref{sec:SpinSum}.

 We henceforth drop the gauge-averaging $SU(2)$ integrals in \eqref{eq:ThiemannState} because in the following sections we will consider the boundary states in contraction with a spinfoam amplitude and so the $\slc$ integrals in the vertex amplitude will automatically implement gauge invariance at the nodes. 
 
 The first simplification we make is to consider states where all semiclassicality parameters $t_\ell$ are set equal
\begin{equation} 
  t_\ell = t \ \forall \ell\in\Gamma
\end{equation}
The semiclassicality parameter $t$ controls the spread of the gaussians over the spins and is thus set to be inversly proportional to a typical \emph{macroscopic} area $A$ of the triangulation
\begin{align} \label{eq:tDef}
	t = \left(\frac{l_p^2}{A}\right)^{n}\quad\text{with } n\in [0, 2],
\end{align}
The multiplication by the Planck area $l_P^2$ is so that $t$ is indeed dimensionless. The reasoning for the restriction of the values of $n$ to $0,1,2$ is explained below. Since by assumption
\begin{equation} \label{eq:macroscopicArea}
A \gg l_p^2,
\end{equation}
we have that 
\begin{equation} \label{eq:larget}
	t \ll 1.
\end{equation}

When
\begin{align}
	\eta_\ell \gg 1 \ \forall \ell\in\Gamma,
\end{align}
the states \eqref{eq:ThiemannState} are proportional to \cite{bianchi_coherent_2010}
\begin{equation}\label{eq:ExtrinsicStates}
	\Psi_{\Gamma, H_\ell}^{t}(h_\ell) \propto \sum_{\{j_\ell \}} \prod_\ell d_{j_\ell} \e^{-\left(j_\ell - \omega_\ell\right)^2 t \,+\, i \gamma \zeta_\ell j_\ell}\, \Phi_{\Gamma, j_\ell, \vec{n}_{s(\ell)}, \vec{n}_{t(\ell)}}(h_\ell),
\end{equation}
where we dropped a multiplicative factor $\prod_\ell \exp((\eta_\ell-t)^2/4t)$ and defined the \emph{area data}
\begin{equation} \label{eq:omegaWithEta}
	\omega_\ell := \frac{\eta_\ell-t}{2t} \approx \frac{\eta_\ell}{2t} .
\end{equation} 
The states $\Phi_{\Gamma, j_\ell, \vec{n}_{s(\ell)}, \vec{n}_{t(\ell)}}(h_\ell)$ are given by
\begin{equation} \label{eq:LSgroup}
\psi_{\Gamma, j_\ell, \vec{n}_{s(\ell)}, \vec{n}_{t(\ell)}}(h_\ell) = \sum_{m_s, m_t} D^{j_\ell}_{j_\ell m_t}(n_{t(\ell)}^\dagger)\; D^{j_\ell}_{m_t m_s}(h_\ell)\; D^{j_\ell}_{m_s j_\ell}(n_{s(\ell)}).
\end{equation}
 The gauged averaged version of the above are the Livine-Speziale coherent states \cite{LS_State} also known as intrinsic coherent states \cite{Rovelli_New_Book}.
 
From the above we see that the large $\eta$ limit of Thiemann's states parametrized in the twisted geometry parametrization indeed corresponds to a large area limit: large $\eta$ implies large $\omega$ from \eqref{eq:larget} and \eqref{eq:omegaWithEta}, with omega admitting a direct interpretation as the macroscopic area on which spins are peaked. The states \eqref{eq:ExtrinsicStates} are peaked on $j_\ell = \omega_\ell$ due to the Gaussian weight factors, which have a spread
\begin{align}
	\sigma := \frac{1}{\sqrt{2t}} \gg 1.
\end{align}
In particular, the expectation values $A_\ell$ of the area operator on these states are given by
\begin{equation}
	A_\ell \approx \gamma\, l_p^2\, \omega_\ell,
\end{equation}
the parameters $\omega_\ell$ (and consequently the parameters $\eta_\ell$) are directly related to physical areas $A_\ell$. We are assuming the Immirzi parameter to be of order unit
\begin{equation}
\gamma \sim 1
\end{equation}

By tuning the parameter $t$, it is possible to peak the states on a prescribed intrinsic and extrinsic semiclassical geometry.  For this to be the case, $t$ has to be chosen such that the spreads in the areas $\Delta A_\ell$ and the spreads in the holonomies $\Delta h_\ell$ are much smaller than the expectation values of the corresponding operators. This requirement translates into
\begin{align}
	\Delta A_\ell \sim \frac{l_p^2}{\sqrt{t}}\ll A_\ell \quad \text{and}\quad \Delta h_\ell \sim \sqrt{t} \ll 1 \quad \forall\ell\in\Gamma.
\end{align}
 Combining the two, we obtain the \emph{semiclassicality condition} that will be used in what follows:
\begin{align}\label{eq:SemiClassicalityConditions}
	& l_p^2 \ll \sqrt{t}\, A_\ell \ll A_\ell\ 
\nonumber\\ \nonumber\\	
	\Leftrightarrow \ \ & 1\ll \sqrt{t}\, \omega_\ell \ll \omega_\ell.
\end{align}
The above translate to the condition $n\in (0, 2)$ for the exponent in \eqref{eq:tDef}.

In summary, the above reflect a physical setup where there is a single typical area scale. This is for instance the case in the transition of a black to a white hole in spherical symmetry, where the relevant area scale is given by $\frac{m^2}{m_P} l_P ^2$, where $m$ is the mass of the hole and $m_P$ is the Planck mass, as in for instance \cite{ChristodoulouCharacteristicTimeScales2018}. In a homogeneous cosmological setup, the relevant area scale would be given by the squared of the area factor. Recall that the bounce in Loop Quantum Cosmology\cite{Banerjee:2011qu,Ashtekar:2013hs} occurs when the typical area is still macroscopic; i.e.\! while we are still in the large areas regime of equation \eqref{eq:macroscopicArea}.

\section{The Path Integral Representation of The Lorentzian EPRL Amplitude} \label{Subsection_PathIntegralRep}

In this section we give a different derivation of the path integral representation of the EPRL amplitude discovered in \cite{Han_Krajewski}. In that work, the authors employed group theoretical methods which allowed them to give a path-integral representation for the case of a 2-complex without boundary that precisely captures the number of degrees of freedom, without the need of introducing auxiliary spinorial variables such as those appearing in the coherent state representation. That has the advantage of rendering the critical point equations in the asymptotic analysis particularly transparent. However, because the method used in \cite{Han_Krajewski} for the derivation differs significantly from the coherent state representation techniques it becomes difficult to combine it with the analysis carried out in \cite{Han_2011}, which is what we want to do here. The derivation of the representation in  \cite{Han_Krajewski} that we give here uses similar techniques as in \cite{Han_2011}. This then allows us to extend the representation of \cite{Han_Krajewski} to two-complexes with boundary. The contraction of the EPRL amplitude with boundary coherent states (that is, performing the spin sum) then becomes straightforward by combining the results of \cite{Han_Krajewski} and \cite{Han_2011}. This is done in Section \ref{sec:decayingAmplitudesCalc}.

First we fix notation and terminology. We will consider a topological two-complex $\mathcal{C}$ with a non-empty boundary $\Gamma:=\partial\mathcal{C}\neq\emptyset$  and bulk $\mathcal{B}:=\mathcal{C}\backslash \Gamma$. The two-complex is assumed to be dual to a four dimensional simplicial triangulation and consists of a collection of five-valent vertices $\ve$ connected by edges $\ed$ which in turn bound faces $\fa$. All vertices belong to the bulk $\mathcal{B}$, but some of the edges which emanate from a vertex intersect the boundary and therefore terminate at a node $\no$. Nodes are four-valent and connected by links $\ell$. Vertices and edges are considered to be part of the bulk structure (also referred to as one-skeleton) while nodes and links constitute the boundary graph. Faces are said to be \emph{bulk faces} when they are bounded by vertices and edges and we also write $\fa\in\mathcal{B}$. If a face is bounded by vertices, edges, nodes and links it is said to be a \emph{boundary face} and we write $\fa\in\Gamma$ with a slight abuse of notation ($\Gamma$ is a graph, it has no faces, it is a boundary link of the face $\fa$ that belongs to $\Gamma$).

\begin{figure}[ht]
	\centering
    \subfigure[Bulk]{\includegraphics[width=0.45\textwidth]{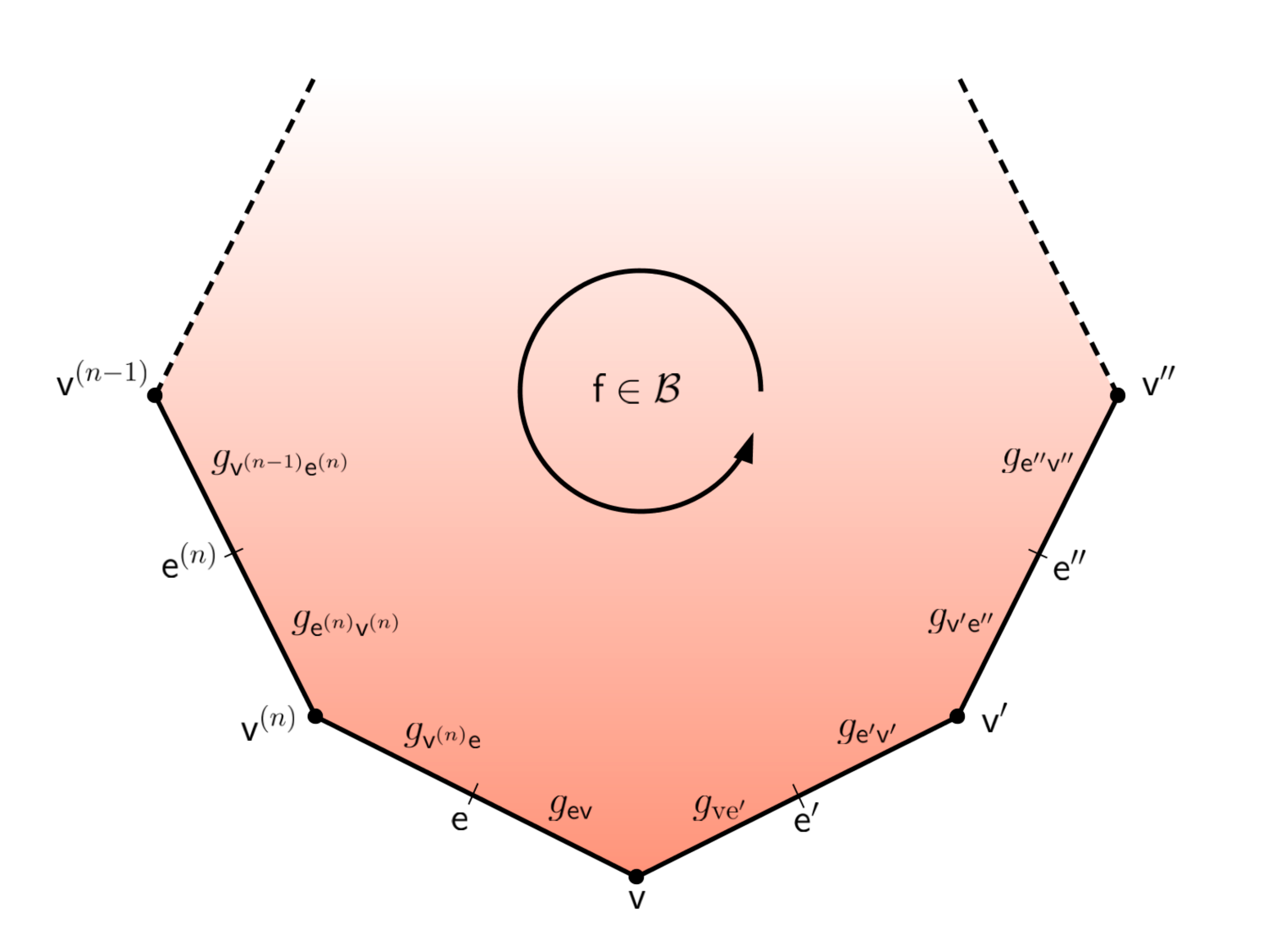}}
    \subfigure[Boundary]{\includegraphics[width=0.45\textwidth]{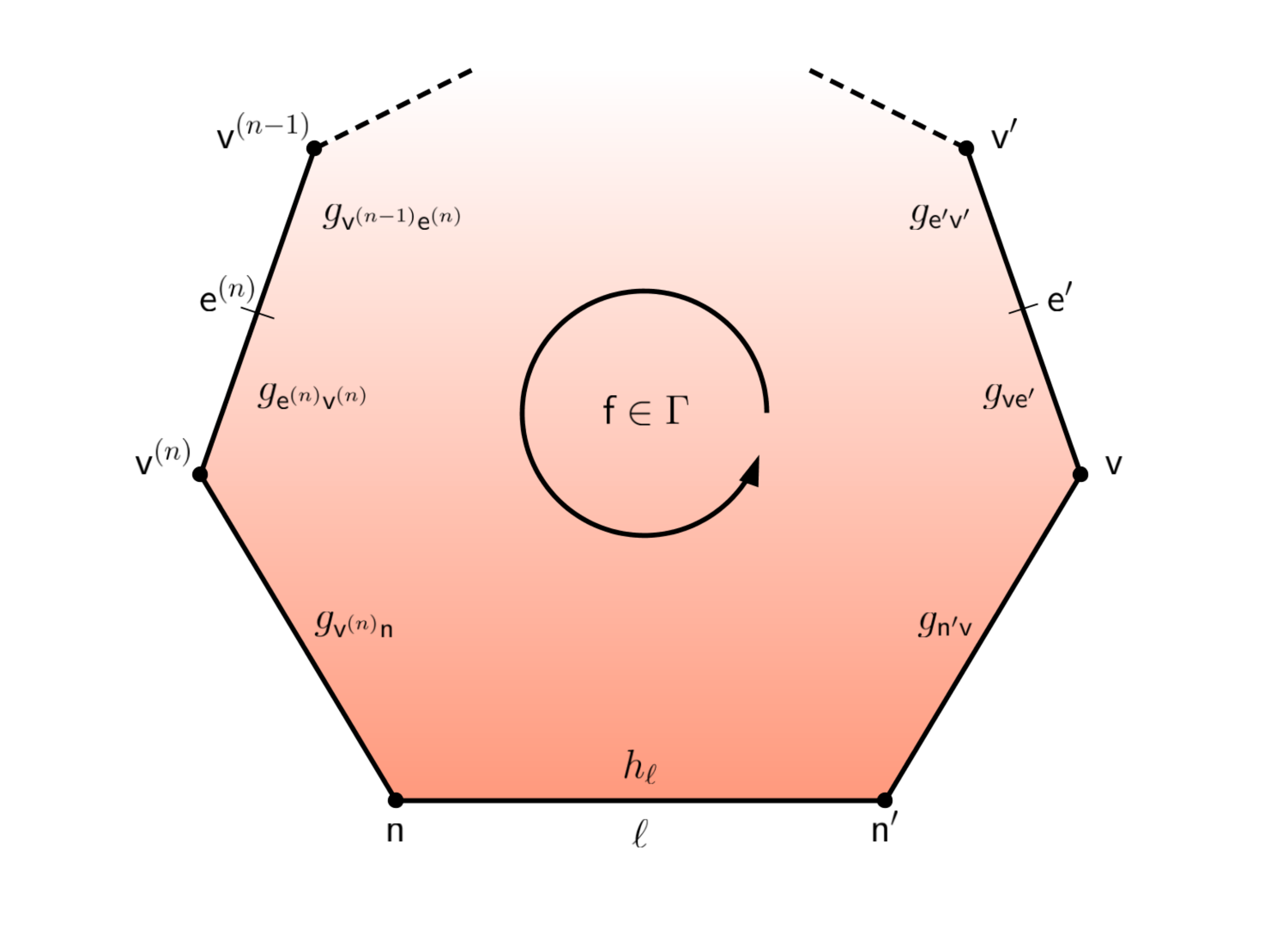}}
    \caption{Notation and Conventions.}
    \label{fig:Notation&Convention}
\end{figure}

The EPRL amplitude is a map $W_\mathcal{C}:\mathcal{H}_\Gamma\to\mathbb{C}$ defined on the two-complex $\mathcal{C}$, which associates complex numbers to states from the boundary Hilbert space $\mathcal{H}_\Gamma = L^2\left[SU(2)^L/SU(2)^N\right]$. A group element $g_{\ve\ed}\in\slc$ is associated to every half edge in the bulk (see Figure \ref{fig:Notation&Convention}) and by convention we set $g_{\ve\ed} = g_{\ed\ve}^{-1}$ . If an edge originating from $\ve$ terminates at a node $\no$, then it is not split in two and the single group element is associated to it is denoted as $g_{\ve\no}\in\slc$ . Links carry $SU(2)$ group elements $h_\ell$ and all faces, whether they are boundary or bulk faces, are colored by a half-integer spin $j_\fa > 0$. Moreover, all faces carry an orientation which in turn induces an orientation on the edges and links (see Figure \ref{fig:Notation&Convention}). In particular, the face orientation induces the notion of ingoing and outgoing group elements. An element of the form $g_{\ed\ve}$ sits on the half edge $\ed$ which enters the vertex $\ve$ and is called ingoing while the element $g_{\ve\ed'}$ sits on the half edge $\ed'$ which exits the vertex $\ve$ and is called outgoing.

The amplitude $W_\mathcal{C}$ can be written as a product of face amplitudes, associated to every face of the two-complex. For bulk faces, the face amplitude $A_\fa$ is constructed as follows: At every vertex $\ve$ we build the product of ingoing group element $g_{\ed\ve}$ and outgoing group element $g_{\ve\ed'}$, i.e. $g_{\ed\ve}g_{\ve\ed'}$. Every such product is multiplied from the left by $Y^\dagger$ and from the right by $Y$, yielding a product of the form $Y^\dagger g_{\ed\ve}g_{\ve\ed'} Y$ at every vertex. The unitary injection $Y$ ($Y$-map) will be defined precisely below. These terms combine as we go around the face, and the face amplitude is defined as
\begin{align}\label{eq:FaceAmplitudes}
	A_\fa := \sum_{j_\fa}d_{j_\fa} \,  \Tr_{j_\fa}\left[\prod_{\ve\in\fa} Y^{\dagger} g^{-1}_{\ve\ed}\,g_{\ve\ed'} Y\right]:= \sum_{j_\fa} d_{j_\fa} \Tr_{j_\fa}\left[ Y^\dagger g_{\ed\ve}g_{\ve\ed'} Y Y^\dagger g_{\ed'\ve'}g_{\ve'\ed''} Y\dots Y^\dagger g_{\ed^{(n)}\ve^{(n)}}g_{\ve^{(n)}\ed} Y \right] \quad\text{for }\fa\in\mathcal{B} ,
\end{align}
where the summation in $j_\fa$ runs over $\frac{1}{2}\mathbb{N}\backslash\{0\}$ in half-integer steps.  The trace is explicitly defined by
\begin{align}\label{eq:def_trace}
	\Tr_{j_\fa}\left[\prod_{\ve\in\fa} Y^{\dagger} g^{-1}_{\ve\ed}\,g_{\ve\ed'}  Y\right] = \sum_{\{m_\ed\}} D^{(\gamma j_\fa, j_\fa)}_{j_\fa m_\ed j_\fa m_\ed'}(g_{\ed\ve}g_{\ve\ed'})  D^{(\gamma j_\fa, j_\fa)}_{j_\fa m_{\ed'} j_\fa m_{\ed''}}(g_{\ed'\ve'}g_{\ve'\ed''}) \dots D^{(\gamma j_\fa, j_\fa)}_{j_\fa m_{\ed^{(n)}} j_\fa m_{\ed}}(g_{\ed^{(n)}\ve^{(n)}}g_{\ve^{(n)}\ed}),
\end{align} 
where $D^{(\gamma j, j)}_{j m j m'}(g)$ are representation matrices of the principal series of unitary irreducible representations of $\slc$, and $\sum_{\{m_\ed\}}$ is a short hand notation for multiple sums (in this case, over all magnetic indices $m_\ed$ appearing in \eqref{eq:def_trace}). The face amplitude $A_\fa(h_\ell)$ for boundary faces is defined analogously, the difference being that edges terminating in nodes are not split into half edges and therefore carry only one $\slc$ group element, and there is an $SU(2)$ group element $h_\ell$ on the link: 
\begin{align}\label{eq:BoundaryFaceAmplitude}
	A_\fa(h_\ell) &:= \sum_{j_\fa}d_{j_\fa} \Tr_{j_\fa}\left[Y^\dagger g_{\ve \no'}^{-1} g_{\ve\ed'} Y \left(\prod_{\ve\in\fa} Y^{\dagger} g^{-1}_{\ve\ed'}\,g_{\ve\ed} Y\right) Y^\dagger g_{\ve^{(n)}\ed^{(n)}}^{-1} g_{\ve^{(n)}\no} Y h_\ell^{-1}\right]  \notag\\
	&= \sum_{j_\fa} d_{j_\fa} \Tr_{j_\fa}\left[Y^\dagger g_{\no'\ve}g_{\ve\ed'} Y Y^\dagger g_{\ed'\ve'}g_{\ve'\ed''} Y \dots Y^\dagger g_{\ed^{(n)}\ve^{(n)}}g_{\ve^{(n)}\no} Y h_\ell^{-1} \right] & \text{for }\fa\in\Gamma.
\end{align}
In the above definition we used the fact that for a 2-complex dual to a simplicial triangulation there is only one link per boundary face.  The amplitude $W_\mathcal{C}(h_\ell)$ associated to the two-complex $\mathcal{C}$ is finally defined as
\begin{align}\label{eq:AmplitudeMap}
	W_\mathcal{C}(h_\ell) &:= \mathcal{N}\int_{\slc}\left(\prod_{\ve}\dd \arc{g}_{\ve\ed}\right)\left(\prod_{\fa\in\mathcal{B}} A_\fa\right)\left(\prod_{\fa\in\Gamma} A_\fa(h_\ell)\right),
\end{align}
where $\mathcal{N}$ is an arbitrary normalization constant. This constant is finite when any one of the five $\slc$ integrations at each vertex is dropped \cite{Barrett_Crane}. This is indicated by the notation $\dd\arc{g}_{\ve\ed}$, which is the product of four $\slc$ Haar measures, explicitly defined as
\begin{align}
	\dd g = \frac{\dd\beta \,\dd\overline\beta \,\dd\gamma \,\dd\overline\gamma \, \dd\delta \,\dd\overline\delta }{|\delta|^2}\quad\text{for}\quad g=\begin{pmatrix}
		\alpha & \beta \\
		\gamma & \delta 
	\end{pmatrix}\in\slc.
\end{align}

\subsection{A different derivation of the Krajewski-Han representation for a Two-Complex without Boundary}
To recast the EPRL amplitude \eqref{eq:AmplitudeMap} in a path integral form, we work in a representation of the principal series of $\slc$ and its subgroup $SU(2)$ on the space of homogeneous functions $\mathcal{H}^{(k, p)}$ in two complex variables $\textbf{z}=(z^0, z^1)^{\transpose}\in\mathbb{C}^2$. A self-contained review of the $\slc$ and $SU(2)$ representation theory on this space is given in Appendix \ref{appendix:App_RepTheory}. Here, we recall only what is necessary for the calculations that follow. The unitary, irreducible, infinite dimensional representations of the principal series of $\slc$ on $\mathcal{H}^{(k, p)}$ are labeled by two parameters, $(k, p)\in\mathbb{R}\times\frac{1}{2}\mathbb{Z}$. In terms of these two parameters,  the functions $F\in\mathcal{H}^{(k, p)}$ satisfy the homogeneity property
\begin{align}
	F(\lambda \textbf{z}) = \lambda^{i k + p -1}\,\overline{\lambda}^{i k - p -1}F(\textbf{z})\quad \forall\lambda\in \mathbb{C}\backslash\{0\}.
\end{align}
The space $\mathcal{H}^{(k, p)}$ decomposes as $\mathcal{H}^{(k, p)}\simeq\bigoplus_{j=\vert p\vert}^{\infty}\mathcal{V}^j$, where $\mathcal{V}^j$ is the space of homogeneous polynomials of degree $2j$ in two complex variables. The $Y$-map provides us with a unitary injection
\begin{align}
	Y:\mathcal{V}^j\rightarrow \mathcal{H}^{(\gamma j, j)}\quad,\quad f(\textbf{z}) \mapsto F(\textbf{z})=\innerp{\textbf{z}}{\textbf{z}}^{i\gamma j - j-1}f(\textbf{z})\quad\forall f\in\mathcal V^j,
\end{align}
where $\innerp{\textbf x}{\textbf y} = \overline x^0 y^0 + \overline x^1 y^1$ is the $SU(2)$ (but not $\slc$) invariant inner product on $\mathbb C^2$. The $Y$-map allows us to easily determine a basis of $\mathcal{H}^{(k, p)}$. As shown in Appendix \ref{appendix:App_RepTheory}, the space $\mathcal V^j$ is spanned by the basis polynomials 
\begin{align}
	P_m^j(\textbf{z}) = \left[\frac{(2j)!}{(j+m)!(j-m)!}\right]^{\frac{1}{2}}z_0^{j+m}\,z_1^{j-m},\quad m\in\{-j, ...,j\}
\end{align}
and acting with the $Y$-map on these basis elements yields
\begin{align}\label{eq:SLPolynomials}
	\phi_{m}^{(\gamma j, j)}(\textbf{z}):= Y\rhd P_m^j(\textbf{z}) = \sqrt{\frac{d_j}{\pi}} \innerp{\textbf{z}}{\textbf{z}}^{i \gamma j -j-1} P_m^j(\textbf{z}),
\end{align}
which is a basis for $\mathcal{H}^{(k, p)}$. The basis $\phi_{m}^{(\gamma j, j)}$ is orthonormal with respect to the inner product
\begin{align}
	\langle f, g \rangle := \int_{\CP}\dd \Omega \, \overline{f(\textbf{z})}\, g(\textbf{z}),\quad\forall f, g\in\mathcal H^{(\gamma j, j)},
\end{align}
where $\dd \Omega = \frac{i}{2}(z^0\dd z^1 - z^1\dd z^0)\wedge(\overline{z}^0\dd \overline{z}^1 - \overline{z}^1\dd \overline{z}^0)$ is a homogeneous and $\slc$ invariant measure on $\mathbb{C}^2\backslash\{0\}\simeq\CP$. By virtue of this inner product, the $\slc$ representation matrices can be written as
\begin{align}\label{eq:SL2CWigner}
	D_{j\,m\,j\,m'}^{(\gamma j, j)}(g) \equiv \bra{j m}Y^\dagger g Y \ket{j m'} = \int_{\CP}\dd \Omega\, \overline{\phi_{m}^{(\gamma j, j)}(\textbf{z})}\, \phi_{m'}^{(\gamma j, j)}(g^\transpose\textbf{z}).
\end{align}
If $g$ lies in the $SU(2)$ subgroup, the usual Wigner $D$-matrices are recovered, see \cite{RuhlBook}. Equation \eqref{eq:SL2CWigner} is crucial in what follows since it is the key to rewrite the trace of the bulk face amplitude \eqref{eq:FaceAmplitudes} in the representation found in \cite{Han_Krajewski}. From the definition of the trace \eqref{eq:def_trace} together with \eqref{eq:SL2CWigner} it follows that
\begin{align}\label{eq:ExpandedTrace}
	\Tr_{j_\fa}\left[\prod_{\ve\in\fa} Y^{\dagger} g^{-1}_{\ve\ed}\,g_{\ve\ed'}  Y\right] &= \sum_{\{m_\ed\}}\prod_{\ve\in\fa}D_{j_\fa\,m_\ed\, j_\fa\, m_{\ed'}}^{(\gamma j_\fa, j_\fa)}(g_{\ve\ed}^{-1} g_{\ve\ed'})\notag\\
	&= \sum_{\{m_\ed\}} \prod_{\ve\in\fa}\int_{\CP}\dd\Omega_{\ve\fa}\,\overline{\phi_{m_\ed}^{(\gamma j_\fa, j_\fa)}(\textbf{z}_{\ve\fa})}\,\phi_{m_{\ed'}}^{(\gamma j_\fa, j_\fa)}(g_{\ve\ed'}^\transpose\, (g_{\ve\ed}^{-1})^\transpose\,\textbf{z}_{\ve\fa})\notag\\
	&= \sum_{\{m_\ed\}} \prod_{\ve\in\fa}\int_{\CP}\dd\Omega_{\ve\fa}\,\overline{\phi_{m_\ed}^{(\gamma j_\fa, j_\fa)}(g_{\ve\ed}^\transpose\,\textbf{z}_{\ve\fa})}\,\phi_{m_{\ed'}}^{(\gamma j_\fa, j_\fa)}(g_{\ve\ed'}^\transpose\,\textbf{z}_{\ve\fa}).
\end{align}
To get the last line we performed the change of integration variables $\textbf{z}_{\ve\fa}\rightarrow g_{\ve\ed}^\transpose\,\textbf{z}_{\ve\fa}$ and used the $\slc$ invariance of the measure $\dd\Omega_{\ve\fa}$.  Exploiting the fact that the trace \eqref{eq:ExpandedTrace} appears under an integral with an $\slc$ Haar measure in \eqref{eq:AmplitudeMap}, we perform the replacement $g_{\ve\ed}\rightarrow \overline{g}_{\ve\ed}$ on all group variables. We define spinorial variables associated to vertices and half edges of a given face:
\begin{align}
	\textbf{Z}_{\ve\ed\fa} := g_{\ve\ed}^{\dagger}\,\textbf{z}_{\ve\fa}\quad,\quad \textbf{Z}_{\ve\ed'\fa} := g_{\ve\ed'}^{\dagger}\,\textbf{z}_{\ve\fa}.
\end{align}
Using the explicit expression \eqref{eq:SLPolynomials} for the basis functions $\phi_m^{(\gamma j, j)}$, the trace is brought to the form
\begin{align}\label{eq:TraceNewRep}
	\Tr_{j_\fa}\left[\prod_{\ve\in\fa} Y^{\dagger} g^{-1}_{\ve\ed}\,g_{\ve\ed'}  Y\right] &= \sum_{\{m_\ed\}} \prod_{\ve\in\fa}\int_{\CP}\dd\Omega_{\ve\fa}\,\overline{\phi_{m_\ed}^{(\gamma j_\fa, j_\fa)}(\textbf{Z}_{\ve\ed\fa})}\,\phi_{m_{\ed'}}^{(\gamma j_\fa, j_\fa)}(\textbf{Z}_{\ve\ed'\fa}) \notag\\
	&=\sum_{\{m_\ed\}}\prod_{\ve\in\fa}\frac{d_{j_\fa}}{\pi}\int_{\CP}\dd\Omega_{\ve\fa}\,\innerp{\textbf{Z}_{\ve\ed\fa}}{\textbf{Z}_{\ve\ed\fa}}^{-i\gamma j_\fa-j_\fa-1}\innerp{\textbf{Z}_{\ve\ed'\fa}}{\textbf{Z}_{\ve\ed'\fa}}^{i\gamma j_\fa-j_\fa-1}P_{m_{\ed'}}^{j_\fa}\left(\textbf{Z}_{\ve\ed'\fa}\right)\,P_{m_\ed}^{j_\fa}\left(\overline{\textbf{Z}}_{\ve\ed\fa}\right).
\end{align}
In the above expression, the spinorial inner products do not depend on any magnetic indices $m_\ed$. Hence, the sums only extend over the $SU(2)$ basis polynomials $P^j_m$. There are two such polynomials per edge $\ed$ which carry the same magnetic index $m_\ed$ (as there are two $\textbf Z$ spinors per edge, but pertaining to different vertices). This follows from the contraction pattern in \eqref{eq:def_trace}. Consequently, the sum $\sum_{\{m_\ed\}}$ decomposes into a certain number\footnote{The number of sums is equal to the number of edges which constitute the face.} of single, independent sums of the form
\begin{align}\label{eq:SumOverMagInd}
	\sum_{\vert m_{\ed'}\vert \leq j_\fa}P_{m_{\ed'}}^{j_\fa}\left(\textbf{Z}_{\ve\ed'\fa}\right)\, P_{m_{\ed'}}^{j_\fa}\left(\overline{\textbf{Z}}_{\ve'\ed'\fa}\right) &= \sum_{\vert m_{\ed'}\vert \leq j_\fa}\frac{(2j_\fa)!}{(j_\fa+m_{\ed'})!(j_\fa-m_{\ed'})!}\left(\overline{Z}^0_{\ve'\ed'\fa}\,Z^0_{\ve\ed'\fa}\right)^{j_\fa+m_{\ed'}}\,\left(\overline{Z}^1_{\ve'\ed'\fa}\,Z^1_{\ve\ed'\fa}\right)^{j_\fa-m_{\ed'}} \notag\\
	&=\sum_{s=0}^{2j_\fa}\binom{2j_\fa}{s}\left(\overline{Z}^0_{\ve'\ed'\fa}\,Z^0_{\ve\ed'\fa}\right)^s \left(\overline{Z}^1_{\ve'\ed'\fa}\,Z^1_{\ve\ed'\fa}\right)^{2j_\fa-s}\notag\\
	&=\left(\overline{Z}^0_{\ve'\ed'\fa}\,Z^0_{\ve\ed'\fa} + \overline{Z}^1_{\ve'\ed'\fa}\,Z^1_{\ve\ed'\fa}\right)^{2j_\fa} =\innerp{\textbf{Z}_{\ve'\ed'\fa}}{\textbf{Z}_{\ve\ed'\fa}}^{2j_\fa}.
\end{align}

In the first line we use the definition \eqref{eq:SLPolynomials} of $P_{m_{\ed'}}^{j_\fa}$ and in the second line we performed the change of summation variable $s=j_\fa+m_{\ed'}$. The resulting binomial sum is trivial and yields the result on the third line. Plugging \eqref{eq:SumOverMagInd} into \eqref{eq:TraceNewRep} and changing from a product over vertices $\ve\in\fa$ to an equivalent product over edges $\ed\in\fa$ brings the bulk face amplitude into the form
\begin{align}\label{eq:RewrittenBulkFace}
	A_\fa &= \sum_{j_\fa}d_{j_\fa}\prod_{\ed\in\fa}\frac{d_{j_\fa}}{\pi}\int_{\CP}\dd\tilde{\Omega}_{\ve\ed\fa}\,\frac{\innerp{\textbf{Z}_{\ve'\ed'\fa}}{\textbf{Z}_{\ve\ed'\fa}}^{2j_\fa}}{\innerp{\textbf{Z}_{\ve'\ed'\fa}}{\textbf{Z}_{\ve'\ed'\fa}}^{i\gamma j_\fa+j_\fa} \innerp{\textbf{Z}_{\ve\ed'\fa}}{\textbf{Z}_{\ve'\ed'\fa}}^{-i\gamma j_\fa+j_\fa}}\notag\\
	&=\sum_{j_\fa}d_{j_\fa}\prod_{\ed\in\fa}\frac{d_{j_\fa}}{\pi}\int_{\CP}\dd\tilde{\Omega}_{\ve\ed\fa}\,\e^{j_\fa S_\fa\left[g_{\ve\ed}, \textbf{z}_{\ve\fa}\right]}&\forall\fa\in\mathcal B.
\end{align}
As in \cite{Han_Krajewski} and \cite{Han_2011} we introduced the rescaled measure 
\begin{align}\label{eq:RescaledMeasure}
	\dd\tilde{\Omega}_{\ve\ed\fa} & := \frac{\dd \Omega_{\ve\fa}}{\innerp{\textbf{Z}_{\ve\ed\fa}}{\textbf{Z}_{\ve\ed\fa}} \innerp{\textbf{Z}_{\ve\ed'\fa}}{\textbf{Z}_{\ve\ed'\fa}}}
\end{align}
and an ``action'' $S_\fa\left[g_{\ve\ed}, \textbf{z}_{\ve\fa}\right]$ associated to bulk faces: 
\begin{align}\label{eq:BulkAction}
	S_\fa[ g_{\ve\ed}, \textbf{z}_{\ve\fa}] & := \log\frac{\innerp{\textbf{Z}_{\ve'\ed'\fa}}{\textbf{Z}_{\ve\ed'\fa}}^{2}}{\innerp{\textbf{Z}_{\ve\ed\fa}}{\textbf{Z}_{\ve\ed\fa}} \innerp{\textbf{Z}_{\ve\ed'\fa}}{\textbf{Z}_{\ve\ed'\fa}}} + i \gamma \log \frac{\innerp{\textbf{Z}_{\ve\ed'\fa}}{\textbf{Z}_{\ve\ed'\fa}}}{\innerp{\textbf{Z}_{\ve\ed\fa}}{\textbf{Z}_{\ve\ed\fa}}}.
\end{align}
When the two-complex has no boundary, i.e. when $\Gamma = \emptyset$, then the EPRL transition amplitude in its path integral form would read 
\begin{align}
	W_\mathcal{C} &= \mathcal N \int_{\slc}\left(\prod_\ve \dd\arc{g}_{\ve\ed}\right)\prod_{\fa\in\mathcal B} \left(\sum_{j_\fa}d_{j_\fa}\prod_{\ed\in\fa}\frac{d_{j_\fa}}{\pi}\int_{\CP}\dd\tilde{\Omega}_{\ve\ed\fa}\,\e^{j_\fa S_\fa\left[g_{\ve\ed}, \textbf{z}_{\ve\fa}\right]}\right) \notag\\
	&= \mathcal N \sum_{\{j_\fa\}} \int_{\slc}\left(\prod_\ve \dd\arc{g}_{\ve\ed}\right)\left(\prod_{\fa\in\mathcal B} d_{j_\fa}\prod_{\ed\in\fa}\frac{d_{j_\fa}}{\pi}\int_{\CP}\dd\tilde{\Omega}_{\ve\ed\fa}\right)\e^{\sum_{\fa\in\mathcal B}j_\fa S_\fa\left[g_{\ve\ed}, \textbf{z}_{\ve\fa}\right]}.
\end{align}
This is precisely the result first obtained in \cite{Han_Krajewski} by different means.

\subsection{Extension of the Krajewski-Han Path Integral representation to Two-Complexes with Boundary}
We now proceed to generalize the calculation of the previous section to a two-complex with a boundary. To this end, it is necessary to also rewrite the trace in the boundary face amplitude \eqref{eq:BoundaryFaceAmplitude} in terms of functions on $\mathcal H^{(\gamma j, j)}$. From the first line of the definition \eqref{eq:BoundaryFaceAmplitude} it follows that the product over vertices can be treated in the same way as for the bulk face amplitude as none of the group elements lives on an edge which terminates in a node. The only group elements we need to consider here are the first two and the last three in the trace of \eqref{eq:BoundaryFaceAmplitude} (see also Figure \ref{fig:Notation&Convention}). For conciseness, we write $\left(\star\right)$ as placeholder for the product over vertices. We have
\begin{align}\label{eq:BoundaryTraceLong}
	\Tr_{j_\fa} & \left[Y^\dagger g_{\ve \no'}^{-1} g_{\ve\ed'} Y \left(\prod_{\ve\in\fa} Y^{\dagger} g^{-1}_{\ve\ed'}\,g_{\ve\ed} Y\right) Y^\dagger g_{\ve^{(n)}\ed^{(n)}}^{-1} g_{\ve^{(n)}\no} Y h_\ell^{-1}\right]\notag\\
	&= \sum_{\{m_\ed\}} D^{(\gamma j_\fa, j_\fa)}_{j_\fa m_{\no'} j_\fa m_{\ed'}}(g_{\ve\no'}^{-1}g_{\ve\ed'}) \left(\star\right) D^{(\gamma j_\fa, j_\fa)}_{j_\fa m_{\ed^{(n)}} j_\fa m_{\no}}(g_{\ve^{(n)}\ed^{(n)}}^{-1}g_{\ve^{(n)}\no}) D^{j_\fa}_{m_\no m_{\no'}}(h^{-1}_\ell)\notag\\
	&= \sum_{\{m_\ed\}}\int_{\CP}\dd\Omega_{\ve\fa}\,\overline{\phi^{(\gamma j_\fa, j_\fa)}_{m_{\no'}}(\textbf z_{\ve\fa})}\phi^{(\gamma j_\fa, j_\fa)}_{m_{\ed'}}(g_{\ve\ed'}^\transpose(g_{\ve\no'}^{-1})^\transpose\textbf z_{\ve\fa}) \left(\star\right) \int_{\CP}\dd\Omega_{\ve^{(n)}\fa}\,\overline{\phi^{(\gamma j_\fa, j_\fa)}_{m_{\ed^{(n)}}}(\textbf z_{\ve^{(n)}\fa})}\phi^{(\gamma j_\fa, j_\fa)}_{m_{\no}}(g_{\ve^{(n)}\no}^\transpose(g_{\ve^{(n)}\ed^{(n)}}^{-1})^\transpose\textbf z_{\ve^{(n)}\fa}) \times\notag\\
	&\quad\times \int_{\CP}\dd\Omega_\ell\, \overline{\phi^{(\gamma j_\fa, j_\fa)}_{m_{\no}}(\textbf z_\ell)}   \phi^{(\gamma j_\fa, j_\fa)}_{m_{\no'}}((h^{-1}_\ell)^\transpose\textbf z_\ell)\notag\\
	&= \sum_{\{m_\ed\}}\int_{\CP}\dd\Omega_{\ve\fa}\,\overline{\phi^{(\gamma j_\fa, j_\fa)}_{m_{\no'}}(\textbf Z_{\ve\no'\fa})}\phi^{(\gamma j_\fa, j_\fa)}_{m_{\ed'}}(\textbf Z_{\ve\ed'\fa}) \left(\star\right) \int_{\CP}\dd\Omega_{\ve^{(n)}\fa}\,\overline{\phi^{(\gamma j_\fa, j_\fa)}_{m_{\ed^{(n)}}}(\textbf Z_{\ve^{(n)}\ed^{(n)}\fa})}\phi^{(\gamma j_\fa, j_\fa)}_{m_{\no}}(\textbf Z_{\ve^{(n)}\no\fa}) \times\notag\\
	&\quad\times \int_{\CP}\dd\Omega_\ell\, \overline{\phi^{(\gamma j_\fa, j_\fa)}_{m_{\no}}(h^\transpose_\ell \textbf z_\ell)}   \phi^{(\gamma j_\fa, j_\fa)}_{m_{\no'}}(\textbf z_\ell).
\end{align}
To get the last line we exploited again the $\slc$ invariance of the measures and performed the same change of integration variables as before. That is, we introduce the following spinorial variables associated to the two edges which terminate in the nodes $\no$, $\no'$:
\begin{align}
	\textbf{Z}_{\ve\no'\fa} := g^\dagger_{\ve\no'} \textbf z_{\ve\fa}\quad,\quad \textbf Z_{\ve^{(n)}\no\fa} := g^\dagger_{\ve^{(n)}\no}\textbf z_{\ve^{(n)}\fa},
\end{align}
Next, we collect only the relevant terms in \eqref{eq:BoundaryTraceLong} and compute 
\begin{align}
	\sum_{m_{\no}, m_{\no'}} &\overline{\phi^{(\gamma j_\fa, j_\fa)}_{m_{\no'}}(\textbf Z_{\ve\no'\fa})}\,    \phi^{(\gamma j_\fa, j_\fa)}_{m_{\no}}(\textbf Z_{\ve^{(n)}\no\fa})\,    \overline{\phi^{(\gamma j_\fa, j_\fa)}_{m_{\no}}(h^\transpose_\ell \textbf z_\ell)}\,   \phi^{(\gamma j_\fa, j_\fa)}_{m_{\no'}}(\textbf z_\ell)\notag\\
	&=\innerp{\textbf Z_{\ve\no'\fa}}{\textbf Z_{\ve\no'\fa}}^{-i \gamma j_\fa - j_\fa - 1} \innerp{\textbf Z_{\ve^{(n)}\no\fa}}{\textbf Z_{\ve^{(n)}\no\fa}}^{i \gamma j_\fa - j_\fa - 1}\innerp{\textbf z_\ell}{\textbf z_\ell}^{-2(j_\fa+1)}\sum_{m_{\no}} P^{j_\fa}_{m_\no}(h^\dagger_\ell \overline{\textbf{z}}_\ell) P^{j_\fa}_{m_\no}(\textbf Z_{\ve^{(n)}\no\fa}) \sum_{m_{\no'}} P^{j_\fa}_{m_\no'}(\overline{\textbf{Z}}_{\ve\no'\fa}) P^{j_\fa}_{m_\no'}(\textbf z_\ell) \notag\\
	&=\innerp{\textbf Z_{\ve\no'\fa}}{\textbf Z_{\ve\no'\fa}}^{-i \gamma j_\fa - j_\fa - 1} \innerp{\textbf Z_{\ve^{(n)}\no\fa}}{\textbf Z_{\ve^{(n)}\no\fa}}^{i \gamma j_\fa - j_\fa - 1}\innerp{\textbf z_\ell}{\textbf z_\ell}^{-2(j_\fa+1)} \innerp{h^\transpose_\ell \textbf z_\ell}{\textbf Z_{\ve^{(n)\no\fa}}}^{2 j_\fa} \innerp{\textbf Z_{\ve\no'\fa}}{\textbf z_\ell}^{2 j_\fa}.
\end{align}
Plugging the above result back into \eqref{eq:BoundaryTraceLong} yields
\begin{align}
	\Tr_{j_\fa} & \left[Y^\dagger g_{\ve \no'}^{-1} g_{\ve\ed'} Y \left(\prod_{\ve\in\fa} Y^{\dagger} g^{-1}_{\ve\ed'}\,g_{\ve\ed} Y\right) Y^\dagger g_{\ve^{(n)}\ed^{(n)}}^{-1} g_{\ve^{(n)}\no} Y h_\ell^{-1}\right] \notag\\
	&= \left(\prod_{\ed\in\fa}\frac{d_{j_\fa}}{\pi}\int_{\CP}\dd\tilde{\Omega}_{\ve\ed\fa}\right)\left(\frac{d^3_{j_\fa}}{\pi^3}\int_{(\CP)^3}\dd\tilde{\Omega}_{\no\ell\no'}\right)\,\e^{j_\fa S_\fa\left[g_{\ve\ed}, \textbf{z}_{\ve\fa}\right] + j_\fa B_\ell[g_{\ve\no}, h_\ell, \textbf z_\ell]},
\end{align}
where we have introduced the rescaled $(\CP)^3$ measures 
\begin{align}\label{eq:BndMeasure}
	\dd\tilde{\Omega}_{\no\ell\no'} &:= \frac{\dd \Omega_{\ve^{(n)}\fa}}{\innerp{\textbf Z_{\ve^{(n)}\no\fa}}{\textbf Z_{\ve^{(n)}\no\fa}}}\frac{\dd\Omega_\ell}{\innerp{\textbf z_\ell}{\textbf z_\ell}^2}\,\frac{\dd \Omega_{\ve\fa}}{\innerp{\textbf Z_{\ve\no'\fa}}{\textbf Z_{\ve\no'\fa}}}
\end{align}
associated to the vertices attached to the nodes $\no$, $\no'$ and to the link $\ell$. Moreover, the action $S_\fa[g_{\ve\ed}, \textbf z_{\ve\fa}]$ is defined as in \eqref{eq:BulkAction}, except that the sum over edges excludes the two edges attached to the nodes. These edges are accounted for in the newly defined boundary face action
\begin{align}\label{eq:BoundaryFaceAction}
	B_\ell[g_{\ve\no}, h_\ell, \textbf z_\ell] &:= \log \frac{\innerp{\textbf{Z}_{\ve\no'\fa}}{\textbf z_\ell}^2}{\innerp{\textbf Z_{\ve\no'\fa}}{\textbf Z_{\ve\no'\fa}} \innerp{\textbf z_\ell}{\textbf z_\ell}} + \log \frac{\innerp{h^\transpose_\ell \textbf z_\ell}{\textbf Z_{\ve^{(n)}\no\fa}}^2}{\innerp{\textbf z_\ell}{\textbf z_\ell}\innerp{\textbf Z_{\ve^{(n)}\no\fa}}{\textbf Z_{\ve^{(n)}\no\fa}}} + i\gamma  \log\frac{\innerp{\textbf Z_{\ve^{(n)}\no\fa}}{\textbf Z_{\ve^{(n)}\no\fa}}}{\innerp{\textbf Z_{\ve\no'\fa}}{\textbf Z_{\ve\no'\fa}}}.
\end{align}
The full EPRL amplitude on a two-complex with boundary in its path integral form is finally given by 
\begin{align}\label{eq:PathIntegral}
	W_\mathcal{C}(h_\ell) = \mathcal N \sum_{\{j_\fa\}} \int_{\slc}\left(\prod_\ve \dd\arc{g}_{\ve\ed}\right)\left(\prod_{\fa\in\mathcal C} d_{j_\fa}\prod_{\ed\in\fa}\frac{d_{j_\fa}}{\pi}\int_{\CP}\dd\tilde{\Omega}_{\ve\ed\fa}\right)\left(\prod_{\ell\in\Gamma}\frac{d^3_{j_\fa}}{\pi^3}\int_{(\CP)^3}\dd\tilde{\Omega}_{\no\ell\no'}\right)\e^{\sum_{\fa\in\mathcal C}j_\fa S_\fa + \sum_{\ell\in\Gamma}j_\fa B_\ell}.
\end{align}
The action term
$B_\ell$ seems oddly asymmetric due to the presence of the $h^\transpose_\ell$ element. This can in principle be remedied by arbitrarily splitting $h_\ell$ in a product of two $SU(2)$ elements. This amounts to splitting the link into half links and makes \eqref{eq:BoundaryFaceAction} appear more symmetric. As we see below, this splitting happens naturally when the amplitude \eqref{eq:PathIntegral} is contracted with coherent boundary states.

\subsection{The Holomorphic Amplitude}
The spinfoam amplitude \eqref{eq:PathIntegral} depends on $L$ arbitrary $SU(2)$ elements and is therefore really a map from $\cal{L}{SU(2)^L/SU(2)^N}$ to the reals, defined on a two-complex. The number associated to a two-complex is called the transition amplitude, obtained from contracting the amplitude \eqref{eq:PathIntegral} with a boundary state. Below, we use the coherent states \eqref{eq:ThiemannState} discussed in subsection \ref{Section_SemiclassicalResOfId} and consider the contraction
\begin{equation}
	W_\mathcal{C}^{t_\ell}(H_\ell):= \left<W_\mathcal{C}\,\vline\,\Psi^{t_\ell}_{\Gamma, H_\ell}\right>:= \int_{SU(2)^L}\left(\prod_{\ell\in\Gamma}\dd h_\ell\right)\, W_\mathcal{C}(h_\ell) \, \Psi^{t_\ell}_{\Gamma, H_\ell}(h_\ell),
\end{equation}
which is known in the literature as the holomorphic amplitude. We have also dropped the gauge averaging $SU(2)$ integrals from $\Psi^{t_\ell}_{\Gamma, H_\ell}$ as the $\slc$ integrals of $W_\mathcal{C}$ take care of gauge-invariance. 

To compute this transition amplitude we need consider only the boundary face amplitude $A_\fa(h_\ell)$ times the state $\Psi^{t_\ell}_{\Gamma, H_\ell}$ and employ the Peter-Weyl theorem. The relevant part of the computation yields
\begin{align}
	\int_{SU(2)^L}&\left(\prod_{\ell\in\Gamma} \dd h_\ell\right) A_\fa(h_\ell)\Psi^{t_\ell}_{\Gamma, H_\ell}(h_\ell) \notag\\
	&= \sum_{\{j_\ell\}}\prod_{\ell\in\Gamma}d_{j_\ell} \e^{-j_\ell(j_\ell+1) t_\ell}\Tr_{j_\ell}\left[Y^\dagger g_{\ve \no'}^{-1} g_{\ve\ed'} Y \left(\prod_{\ve\in\fa} Y^{\dagger} g^{-1}_{\ve\ed'}\,g_{\ve\ed} Y\right) Y^\dagger g_{\ve^{(n)}\ed^{(n)}}^{-1} g_{\ve^{(n)}\no} Y H_\ell^{-1}\right].
\end{align}
The integration exchanged the arbitrary $h_\ell^{-1}\in SU(2)$ with the group elements $H_\ell^{-1}\in \slc$ given by \eqref{eq:HParam} and completely determined by the boundary data\footnote{This is achieved by analytical continuation of the Wigner D-matrix $D^j_{ab}(H^{-1}_\ell)$ to $\slc$, see \cite{Barrett_2009,RuhlBook}.}. Moreover, the integration gives rise to a $\delta^{j_\fa j_\ell}$, which forces the spins $j_\fa$ colouring the boundary face to be the same as the spins $j_\ell$ appearing in the boundary states and which live on the links.

Rewriting this trace in terms of functions on $\mathcal H^{(\gamma j_\ell, j_\ell)}$ involves the same steps as in the previous subsection. However, before continuing we will make use of an approximation that is pertinent for the physical applications we have in mind. We are interested in boundary states peaked on geometries with large areas, that is, states with $\eta_\ell\gg 1$, for which the highest weight approximation is appropriate \cite{Bianchi:2009ky}
\begin{align}
	D^{j_\ell}_{ab}(H^{-1}_\ell) = D^{j_\ell}(n_{s(\ell)}\e^{(\eta_\ell + i \gamma\zeta_\ell)\frac{\sigma_3}{2}}n^{-1}_{t(\ell)}) = D^{j_\ell}_{a j_\ell}(n_{s(\ell)}) D^{j_\ell}_{j_\ell b}(n^{-1}_{t(\ell)})\,\e^{(\eta_\ell + i \gamma\zeta_\ell)j_\ell}\left(1+\mathcal O(\e^{-\eta_\ell})\right).
\end{align} 
To rewrite the two Wigner matrices of the $SU(2)$ matrices, which essentially split the link into two parts, we make use of \eqref{eq:SLPolynomials} to obtain
\begin{align}\label{eq:ImportantRelation}
	\phi_{\;j}^{(\gamma j, j)}(n^\transpose\textbf{z}) = \sqrt{\frac{d_j}{\pi}}\innerp{\textbf{z}}{\textbf{z}}^{i\gamma j - j-1}\innerp{\overline{\textbf{n}}}{\textbf{z}}^{2j},
\end{align}
where $\textbf n$ is the spinor corresponding to the $SU(2)$ element $n$. This yields
\begin{align}
	D^{j_\ell}_{a j_\ell}(n_{s(\ell)}) &= \frac{d_{j_\ell}}{\pi} \int_{\CP}\dd\Omega_{\no\ell}\,\innerp{\textbf z_{\no\ell}}{\textbf z_{\no\ell}}^{-2(j_\ell+1)}\innerp{\overline{\textbf{n}}_{s(\ell)}}{\textbf z_{\no\ell}}^{2j_\ell} \overline{P^{j_\ell}_a(\textbf z_{\no\ell})}\notag\\
	D^{j_\ell}_{j_\ell b}(n^{-1}_{t(\ell)}) &= \frac{d_{j_\ell}}{\pi} \int_{\CP}\dd\Omega_{\no'\ell}\,\innerp{\textbf z_{\no'\ell}}{\textbf z_{\no'\ell}}^{-2(j_\ell+1)}\innerp{\textbf z_{\no'\ell}}{\overline{\textbf{n}}_{t(\ell)}}^{2j_\ell} P^{j_\ell}_b(\textbf z_{\no'\ell}).
\end{align}
Repeating the same steps as in the previous section one arrives without much effort at 
\begin{align}\label{eq:HolomorphicFaceAmplitude}
	\e^{-j_\ell(j_\ell+1) t_\ell}&\Tr_{j_\ell}\left[Y^\dagger g_{\ve \no'}^{-1} g_{\ve\ed'} Y \left(\prod_{\ve\in\fa} Y^{\dagger} g^{-1}_{\ve\ed'}\,g_{\ve\ed} Y\right) Y^\dagger g_{\ve^{(n)}\ed^{(n)}}^{-1} g_{\ve^{(n)}\no} Y H_\ell^{-1}\right]\notag\\
	&= \e^{\frac{(\eta_\ell-t_\ell)^2}{4t_\ell}}\left(\prod_{\ed\in\fa}\frac{d_{j_\ell}}{\pi}\int_{\CP}\dd\tilde{\Omega}_{\ve\ed\fa}\right)\left(\frac{d^4_{j_\ell}}{\pi^4}\int_{(\CP)^4}\dd\tilde{\Omega}_{s\ell t}\right)\,\e^{j_\ell F_\fa\left[g_{\ve\ed}, \textbf{z}_{\ve\fa}\right] + B_\ell[j_\ell; H_\ell]}\left(1+\mathcal O(\e^{-\eta_\ell})\right),
\end{align}
where we defined
\begin{align}
	\dd\tilde\Omega_{s\ell t} := \frac{\dd\Omega_{\ve^{(n)}\fa}}{\innerp{\textbf{Z}_{\ve^{(n)}\no\fa}}{\textbf{Z}_{\ve^{(n)}\no\fa}}}\,\frac{\dd\Omega_{\no\ell}}{\innerp{\textbf z_{\no\ell}}{\textbf z_{\no\ell}}^2}\,\frac{\dd\Omega_{\no'\ell}}{\innerp{\textbf z_{\no'\ell}}{\textbf z_{\no'\ell}}^2}\,\frac{\dd\Omega_{\ve\fa}}{\innerp{\textbf Z_{\ve\no'\fa}}{\textbf Z_{\ve\no'\fa}}}
\end{align}
and
\begin{align}\label{eq:GWeight}
	F_\ell[g_{\ve\ed}, \textbf z_{\no\ell};\textbf n_{\no(\ell)}] &:= S_\ell[g_{\ve\ed}, \textbf z_{\no\ell}]+\log \frac{\langle \overline{\textbf{n}}_{s(\ell)}\vert \textbf{z}_{\no\ell} \rangle^2 \langle \textbf{z}_{\no'\ell} \vert \overline{\textbf{n}}_{t(\ell)} \rangle^2}{\innerp{\textbf{z}_{\no\ell}}{\textbf{z}_{\no\ell}}^2 \innerp{\textbf{z}_{\no'\ell}}{\textbf{z}_{\no'\ell}}^2} +  \log \frac{\langle \textbf{Z}_{\ve\no'\ell} \vert\textbf{z}_{\no'\ell} \rangle^2 \langle \textbf{z}_{\no\ell}\vert \textbf{Z}_{\ve^{(n)}\no\ell} \rangle^2}{\langle \textbf{Z}_{\ve\no'\ell} \vert \textbf{Z}_{\ve\no'\ell} \rangle \langle \textbf{Z}_{\ve^{(n)}\no\ell} \vert \textbf{Z}_{\ve^{(n)}\no\ell} \rangle}
	+ i \gamma \log \frac{\langle \textbf{Z}_{\ve^{(n)}\no\ell} \vert \textbf{Z}_{\ve^{(n)}\no\ell}\rangle}{\langle \textbf{Z}_{\ve\no'\ell} \vert \textbf{Z}_{\ve\no'\ell} \rangle} \notag\\
	G_\ell[j_\ell; H_\ell]&:= i\gamma j_\ell \zeta_\ell - \left(j_\ell-\omega_\ell(\eta_\ell, t_\ell)\right)^2 t_\ell.
\end{align}
The definition of $\dd\tilde\Omega_{\ve\ed\fa}$ and $S_\ell[g_{\ve\ed}, \textbf z_{\ve\fa}]$ remain the same as in the previous sections and the exponential pre-factor $\exp((\eta_\ell-t_\ell)^2/4t_\ell)$, which only depends on the data $\eta_\ell$ and the parameter $t_\ell$, arises from completing the square such that the Gaussian weight $\exp(-(j_\ell - \omega_\ell)^2 t_\ell)$ appears in \eqref{eq:GWeight}. Absorbing the pre-factor into the normalization $\mathcal N$ of the EPRL amplitude we finally arrive at the holomorphic amplitude in its path integral form: 
\begin{align}\label{eq:HolomorphicAmplitude}
	W^{t_\ell}_\mathcal{C}(H_\ell) = \mathcal N \sum_{\{j_\fa, j_\ell\}} \int_{\slc}\left(\prod_\ve \dd\arc{g}_{\ve\ed}\right)\left(\prod_{\fa\in\mathcal C} d_{j_\fa}\prod_{\ed\in\fa}\frac{d_{j_\fa}}{\pi}\int_{\CP}\dd\tilde{\Omega}_{\ve\ed\fa}\right)\left(\prod_{\ell\in\Gamma}\frac{d^4_{j_\ell}}{\pi^4}\int_{(\CP)^4}\dd\tilde{\Omega}_{s\ell t}\right)\e^{\sum_{\fa\in\mathcal B}j_\fa S_\fa + \sum_{\ell\in\Gamma}(j_\ell F_\ell+G_\ell)}.
\end{align}
This amplitude is the object of main interest in this paper. In the next section, we will give an approximate expression for this amplitude when defined  two-complexes without interior faces.

\section{Approximation of the EPRL Amplitude on Tree-Level Two-Complexes}\label{sec:SpinSum}

In the previous section we derived the Lorentzian EPRL amplitude in the Krajewski-Han path integral representation  \cite{Han_Krajewski} in a formalism suitable for our purposes, and extended it to two-complexes with boundary. The arising boundary terms and the extended amplitude are summarized in \eqref{eq:BndMeasure}, \eqref{eq:BoundaryFaceAction}, \eqref{eq:PathIntegral}.\newline
The large spin asymptotic of the bulk partial amplitude in \cite{Engle:2007wy} have been studied in detail in \cite{Han_Krajewski} using of the coherent state representation of the EPRL amplitude. The results corroborate the ones derived in \cite{Han_2011}. In \cite{Han_2011}, the authors also gave a detailed analysis of the boundary partial amplitude, again using the coherent state representation. With the analysis of the previous section, we can now combine these results to proceed to the coherent state representation of the transition amplitude. This simply amounts to inserting resolutions of the identity in terms of $SU(2)$ coherent states at each bulk face and does not affect the asymptotics of the boundary partial amplitude as performed in \cite{Han_2011}. Therefore, all results obtained in the coherent state representation will carry over to the Krajewski-Han path integral representation.

The above will be used in this section to develop an approximation of the holomorphic EPRL amplitude~\eqref{eq:HolomorphicAmplitude} defined on a special class of two-complexes. On a general two-complex, it is a difficult task to perform the spin-sum analytically while keeping the approximation scheme under control. Attempts in this direction can be found in \cite{Han_2016, Oliveira_2017, Speziale:2016axj}. Another option that has been explored is to use symmetry reduced models \cite{Bahr:2015gxa, Bahr:2016hwc, Bahr:2017ajs}. Here, we only consider what we call tree-level two-complexes $\mathcal T$: these are dual to a four-dimensional simplicial triangulation of spacetime as before, but with the additional restriction that they only have boundary faces $\fa\in\Gamma:=\partial\mathcal T$. That is, there are no faces which lie completely in the bulk. Considering only these two-complexes allows us to carry the out the calculation to the end. This comes from the observation that the subset of extrinsic boundary states \eqref{eq:ExtrinsicStates} which are tuned to satisfy the semiclassicality condition~\eqref{eq:SemiClassicalityConditions} are sharply peaked on spin values $\omega_\ell$, which are taken to correspond to macroscopic classical areas of a discrete boundary geometry. This peakedness manifests itself in the Gaussian weight factors $\exp(-(j_\ell - \omega_\ell)^2 t_\ell)$ present in \eqref{eq:HolomorphicFaceAmplitude}. These weight factors provide a strong regulator for the boundary face amplitude and they allow to truncate the spin-sums over boundary spins while keeping the approximation under control.

The class of is quite restrictive, but, it is relevant for existing studies of possible physical applications for spinfoams. A concrete example of such a two-complex can be found in \cite{christodoulou_realistic_2016,
ChristodoulouCharacteristicTimeScales2018,MariosGeometryTransitionCovariant2018}, where it has been used to model the transition of a black hole into a white hole.  It is of course desirable to consider also bulk faces. This which is beyond the scope of the present work and is left for future analysis.

\subsection{Truncated Spin-Sums, Triangle Inequalities and Semiclassicality}
The holomorphic amplitude in the highest weight approximation defined on a tree-level two-complex is formally given by
\begin{align}
	W^{t_\ell}_\mathcal{T}(H_\ell) = \mathcal N \sum_{\{j_\ell\}} \int_{\slc}\left(\prod_\ve \dd\arc{g}_{\ve\ed}\right)\left(\prod_{\fa\in\Gamma} d_{j_\ell}\prod_{\ed\in\fa}\frac{d_{j_\ell}}{\pi}\int_{\CP}\dd\tilde{\Omega}_{\ve\ed\fa}\right)\left(\prod_{\ell\in\Gamma}\frac{d^4_{j_\ell }}{\pi^4}\int_{(\CP)^4}\dd\tilde{\Omega}_{s\ell t}\right)\e^{\sum_{\fa\in\Gamma}j_\fa F_\ell + \sum_{\ell\in\Gamma}G_\ell}.
\end{align}
In what follows it is not necessary to keep track of all the details given in the precise definitions of the previous sections. In order to make the discussion in this section more concise we drop most of the indices referring to the structure of the two-complex and rewrite the amplitude as 
\begin{align}\label{eq:SimpleAmplitude}
	W^t_\mathcal{T}(H_\ell) = \mathcal N\sum_{\{j_\ell\}\in D^k_{\omega}} \mu_j \e^{-t\sum_\ell (j_\ell-\omega_\ell)^2}\e^{i\gamma\sum_\ell \zeta_\ell j_\ell}\int_{D_{g, \textbf z}}\dd \mu_{g, \Omega} \e^{\sum_\ell j_\ell F_\ell(g,\textbf z;\textbf n_{\ell(\no)})}.
\end{align}
The notation 
\begin{align}
	\int_{D_{g, \textbf z}} \dd\mu_{g, \Omega} := \int_{\slc}\left(\prod_{\ve}\dd \arc{g}_{\ve\ed}\right)\left(\prod_{\fa\in\Gamma}\prod_{\ed\in\fa}\int_{\CP}\dd\tilde\Omega_{\ve\ed\fa} \right)\left(\prod_{\ell\in\Gamma}\int_{(\CP)^4}\dd\tilde\Omega_{s\ell t}\right)
\end{align}
has been introduced to summarize all $\slc$ and $\CP$ integrals while the notation
\begin{align}\label{eq:SummationMeasure}
	\mu_j:=\left(\prod_{\fa\in\Gamma}\prod_{\ed\in\fa} d_{j_\ell} \right)\left(\prod_{\ell\in\Gamma}d^4_{j_\ell}\right),
\end{align}
represents the summation ``measure'', and where irrelevant factors of $\pi$ have been absorbed into the normalization $\cal{N}$. Moreover, the summation over boundary spins is only performed over the domain
\begin{equation}
	D^k_{\omega} := \underset{\ell}{\largetimes}\left\{\floor*{\omega_\ell - \frac{k}{\sqrt{2t}}}, \floor*{\omega_\ell + \frac{k}{\sqrt{2t}}}\right\} \quad \text{with }\quad 0<k\in \mathbb{N}.
\end{equation}
The symbol $\floor*{x}$ denotes the floor function which, in this article, is defined to be the largest \textit{half integer} number equal to or less than $x$. The restriction to the summation domain $D_\omega^k$ implements the truncation of the spin-sum discussed in the introduction of this section.\footnote{That this is a good approximation of the actual sum follows from the fact that the partial amplitude is an oscillating and finite function of the spins. The Gaussian weights therefore strongly dominate. Further justification is provided by the procedure performed in \cite{Han_2016}, where the author introduced a regulator $\sim\e^{-j}$ to study phase transitions in large spin foams. Here, the coherent states naturally provide us with the stronger regulator $\sim\e^{-j^2}$.} The Gaussian weight factors \eqref{eq:GWeight} regulate the spin-sums while the parameter $k$ acts as cut-off. It measures how many standard deviations $\sigma=1/\sqrt{2t}$ the summation moves away from the peak $\omega_\ell$ and is of order unit. 

The main subtlety that needs to be addressed in what follows is that the sums in \eqref{eq:SimpleAmplitude} cannot immediately be treated as independent. The summand vanishes when the triangle inequalities among the spins are not satisfied. More precisely, the summand in \eqref{eq:SimpleAmplitude} vanishes whenever any one of the intertwiner spaces associated to the nodes of the two-complex is of dimension zero. Therefore, in order to treat the sums as independent and exchange them with the integrals, the spin-sums need to be restricted to spin-configurations for which the intertwiner space is always non-trivial. Let us now see that this is not an issue for the set up of this work. 

To implement this requirement and since by assumption the nodes of the two-complex are four-valent we define the set
\begin{align}\label{eq:OmegaGamma}
	D_{\Gamma} &:= \left\{\{j_\ell\} \bigg\vert \dim\text{Inv}_{SU(2)}\left[\overset{4}{\underset{\ell=1}{\bigotimes}} \mathcal{H}_{j_\ell} \right] > 0 \quad\forall \no \in \Gamma\right\} \notag\\
	&\text{ }=\left\{\{j_\ell\} \bigg\vert \min\left(j_1+j_2, j_3+j_4\right) - \max\left(\vert j_1 - j_2\vert, \vert j_3 - j_4\vert\right) + 1 > 0 \quad\forall \no \in \Gamma\right\}.
\end{align}
This is the set of all spin configurations $\{j_\ell\}$ for which the intertwiner spaces over the whole boundary graph $\Gamma$ are non-trivial. To adequately truncate the spin-sums we must now choose the cut-off parameter $k$ such that 
\begin{align}\label{eq:Implication}
	\{j_\ell\}\in D^k_\omega\subseteq D_\Gamma.
\end{align}
To rewrite this condition, it is convenient to split the boundary spins  $j_\ell$ into fixed background contributions $\lambda a_\ell$ and \textit{fluctuations} $s_\ell$, i.e.
\begin{equation}\label{eq:JDecomposition}
	j_\ell = \lambda a_\ell + s_\ell \quad \text{with}\quad \omega_\ell \equiv \lambda a_\ell \quad\text{and}\quad s_\ell \in \CBr*{-\floor*{\frac{k}{\sqrt{2t}}}, \floor*{\frac{k}{\sqrt{2t}}}}\quad\forall\ell\in\Gamma.
\end{equation}
In this decomposition the $a_\ell$'s are assumed to be of order unit in $\lambda$ and $\lambda\gg 1$. Combining \eqref{eq:JDecomposition} with \eqref{eq:Implication} leads to
\begin{align}
	&\lambda\, a_{\text{sum}} - \frac{2k}{\sqrt{2t}} + 1 > \lambda\, a_{\text{diff}} + \frac{2k}{\sqrt{2t}}\notag\\
	 a_\text{sum} := \min&\left(a_1 + a_2, a_3 + a_4\right) \quad\quad a_\text{diff} := \max\left(\vert a_1 - a_2\vert, \vert a_3 - a_4 \vert\right)
\end{align}
which can be rearranged to
\begin{equation}\label{eq:SemiclassicalityCondition}
	\lambda\, \left(a_\text{sum}-a_\text{diff}\right) \sqrt{t} > \frac{4k}{\sqrt{2}}-\sqrt{t} \approx \frac{4k}{\sqrt{2}}
\end{equation}
since $t$ was assumed to be much smaller than unit \eqref{eq:larget}. By the assumptions of this section, the difference $\lambda\,a_\text{diff}$ is negligible compared to the sum $\lambda\,a_{\text{sum}}$ and hence \eqref{eq:SemiclassicalityCondition} is satisfied when the semiclassicality condition \eqref{eq:SemiClassicalityConditions} holds. The semiclassicality condition can also be read as a geometricity condition on the coherent states. It imposes that the intrinsic states \eqref{eq:LSgroup} have spins which are well within the triangle inequalities. This in turn means that the coherent states are composed of a superposition of intrinsic coherent states \eqref{eq:LSgroup} each peaked on a triangulation of a spacelike hypersurface.

Next we turn to the dimension factors $d_{j_\ell}$. From \eqref{eq:SemiclassicalityCondition} and applying the decomposition \eqref{eq:JDecomposition} we get from \eqref{eq:SummationMeasure}
\begin{align}
	\mu_j &=\left(\prod_{\fa\in\Gamma}\prod_{\ed\in\fa} (2 j_\ell+1) \right)\left(\prod_{\ell\in\Gamma}(2 j_\ell+1)^4\right)\approx \left(\prod_{\fa\in\Gamma}\prod_{\ed\in\fa} 2 j_\ell \right)\left(\prod_{\ell\in\Gamma}(2 j_\ell)^4\right)\notag\\
	&= 2^{N_\mathcal{C}}\left(\prod_{\fa\in\Gamma}\prod_{\ed\in\fa} (\lambda a_\ell+s_\ell) \right)\left(\prod_{\ell\in\Gamma}(\lambda a_\ell+s_\ell)^4\right)\notag\\
	&=\left(2 \lambda a_\ell\right)^{N_\mathcal{C}}\left(1+\mathcal{O}\left(\frac{s_\ell}{\lambda a_\ell}\right)\right)
\end{align}
Dropping $\mathcal{O}(s_\ell/\lambda\,a_\ell)$ is justified when $\vert s_\ell\vert \ll \lambda\, a_\ell$ which is equivalent to $\frac{k}{\sqrt{2}} \ll \lambda\, a_\ell\sqrt{t}$. But this again follows from by the semiclassicality condition \eqref{eq:SemiClassicalityConditions}, since $k$ is of order unit and hence we can safely drop the $\mathcal{O}(s_\ell/\lambda\,a_\ell)$ term. 

\subsection{Performing the Spin-Sum}
\label{sec:decayingAmplitudesCalc}
Due to the semiclassicality condition, the spin-sums over the finite summation domain $D^k_\omega$ (with $k$ chosen appropriately) can be treated as independent. After applying the decomposition \eqref{eq:JDecomposition} to the holomorphic amplitude \eqref{eq:SimpleAmplitude}, it can be rewritten as
\begin{align}
	W^t_\Gamma(H_\ell) = \mathcal N \int_{D_{g, \textbf z}} \mu_j \dd \mu_{g, \Omega} \mathcal U(g, \textbf z; t, H_\ell)\e^{\lambda \Sigma(a_\ell,g, \textbf z; \textbf n_{\ell(\no)})}
\end{align}
where
\begin{align}
	\mathcal U(g, \textbf z; t, H_\ell) := \prod_\ell\left(\sum_{s_\ell\in D^k_{\omega}}\e^{-s^2_\ell t + (i\gamma \zeta_\ell + F_\ell(g, \textbf z; \textbf n_{\ell(\no)})) s_\ell}\right) ,  \qquad \Sigma(a_\ell,g, \textbf z; \textbf n_{\ell(\no)}):=\sum_\ell (a_\ell F_\ell(g, \textbf z; \textbf n_{\ell(\no)})+i\gamma\zeta_\ell a_\ell)
\end{align}

The large parameter $\lambda$ only appears linearly in the exponent and the newly defined function $\mathcal U$ is continuous in the variables $g$ and $\textbf z$. Hence, the generalized stationary phase theorem \cite{Hormander} may be applied.
The critical point equations
\begin{align}
	\text{Re} \Sigma(a_\ell,g, \textbf z; \textbf n_{\ell(\no)}) = \delta_g \Sigma(a_\ell,g, \textbf z; \textbf n_{\ell(\no)}) = \delta_{\textbf z} \Sigma(a_\ell,g, \textbf z; \textbf n_{\ell(\no)})
\end{align}
are exactly those of the fixed-spin asymptotics of \cite{Han_2011} and hence their results can directly be used here. The data $H_\ell$ provided by the semiclassical states is either Regge-like, in which case there will be a geometrical critical point corresponding to one of three possible types of simplicial geometries, or there will be no critical point. We may assume the data $(\omega_\ell, n_{\ell(\no)})$ to be Regge-like and moreover we may choose it such that vector geometries are excluded. Also need to mention that there are $2^N$ critical points.

By virtue of the stationary phase theorem we have the following estimation for the amplitude

\begin{equation}\label{my ampl}
    W^t_\mathcal{T}(H_\ell)=N\sum_c \mu_j \lambda^{M^c_\mathcal{C}}\mathcal{H}_c(a_\ell,\textbf n_{\ell(\no)}) \mathcal{U}(g_c,z_c;t,H_l)\e^{\lambda \Sigma(a_\ell,g, \textbf z; \textbf n_{\ell(\no)})}\left(1+\mathcal{O}(\lambda^{-1})\right),
\end{equation}
where $\mathcal{H}_c$ contains the determinant of the Hessian of $\Sigma$. The important point to keep for physical applications is that in the first order approximation, the scale $\lambda$ appears only as an overall scaling factor $\lambda^{M^c_\mathcal{C}}$ and as a linear term in the exponential. In particular, $\mathcal{H}_c$ does not depend on $\lambda$.

We proceed to evaluate $\mathcal{U}$ at the critical point by using 
\begin{equation}
F_\ell(g, \textbf z; \textbf n_{\ell(\no)}) = - i \gamma \, \phi_\ell(s_{c(\ve)}, a_\ell, \textbf n_{\ell(\no)}),
\end{equation}
where $\phi_\ell(s_{c(\ve)}, a_\ell, \textbf n_{\ell(\no)})$ is the Palatini deficit angle. Thus, $\mathcal{U}$ evaluated at $c$ reads 
\begin{equation} \label{eq:tempUtilde}
	\mathcal U(g_c, \textbf z_c; t, H_\ell) = \prod_\ell\left(\sum_{s_\ell\in D^k_{\omega}}\e^{-s^2_\ell t + i\gamma(\zeta_\ell - \phi_\ell(g, \textbf z; \textbf n_{\ell(\no)})) s_\ell}\right)
\end{equation}
Since the phase $i \gamma (\zeta_\ell-\phi_\ell)$  is purely imaginary and independent of $s_\ell$, the sum is dominated by the exponential damping factor $\exp(-s_\ell^2 t)$. It can reasonably be expected that due to this exponential damping the sum converges very fast and that it is therefore a good approximation to remove the cut-off $k$ and sum $s_\ell$ from $-\infty$ to $\infty$ for all $\ell\in\Gamma$. This allows us to get a closed analytic expression for the spin-sums, which approximates them well:
\begin{align}
	\sum_{s_\ell = -\infty}^{\infty}\e^{-s^2_\ell t + i \gamma (\zeta_\ell - \phi_\ell) s_\ell} = 2\sqrt{\frac{\pi}{t}}\e^{-\frac{\gamma^2}{4t}(\zeta_\ell-\phi_\ell)^2}\vartheta_3\left(-\frac{i\pi\gamma(\zeta_\ell-\phi_\ell)}{t}, \e^{-\frac{4\pi^2}{t}}\right),
\end{align}
where
\begin{align}
	\vartheta_3(u, q) := 1 + 2\sum_{n=1}^{\infty}q^{n^2} \cos(2nu)
\end{align}
is the third Jacobi theta function. Hence, 
\begin{equation}
    \mathcal U(g_c, \textbf z_c; t, H_\ell) \approx \prod_\ell 2\sqrt{\frac{\pi}{t}}\e^{-\frac{\gamma^2}{4t}(\zeta_\ell-\phi_\ell)^2}\vartheta_3\left(-\frac{i\pi\gamma(\zeta_\ell-\phi_\ell)}{t}, \e^{-\frac{4\pi^2}{t}}\right).
\end{equation}
Substituting everything to \eqref{my ampl} we obtain
\begin{equation}
    W^t_\mathcal{T}(H_\ell)=\mathcal{N}\sum_c\lambda^N\mu(a)\prod_\ell\left(\e^{-\frac{\gamma^2}{4t}(\zeta_\ell-\phi_\ell)^2}\vartheta_3\left(-\frac{i\pi\gamma(\zeta_\ell-\phi_\ell)}{t}, \e^{-\frac{4\pi^2}{t}}\right)\right)e^{\sum_\ell(-\lambda i\gamma a_\ell \phi_\ell(s_{c(\ve)}, a_\ell, \textbf n_{\ell(\no)}) + i \lambda\gamma \zeta_\ell a_\ell)  }
\end{equation}
The power $N$ is in general a half integer that depends on the rank of the hessian at the critical point and the combinatorics of the 2-complex C. The function $\mu(a)$ includes the summation measure over the spins and the Hessian evaluated at the critical point.

 In Appendix \ref{appendixB} it is explained that for our purposes $\theta_3\approx 1$ can be approximated by unit. Thus, we obtain
\begin{equation}\label{almost final}
     W^t_\mathcal{T}(H_\ell)\approx \mathcal{N}\sum_c\lambda^N\mu(a)\prod_\ell e^{\frac{-{\gamma}^2}{4t}(\zeta_\ell-\phi_\ell)^2+i\gamma(\zeta_\ell-\phi_\ell)\omega_\ell}\left(1+\mathcal{O}(\lambda^{-1})\right )
\end{equation}
The above result can be generalized to include all geometric cases of critical points. Following the same procedure we arrive at

\begin{equation}\label{final}
    W^t_\mathcal{T}(H_\ell)\approx \mathcal{N}\sum_c\lambda^N\mu(a)\prod_\ell e^{\frac{-{\Delta_\ell}^2}{4t}+i\Delta_\ell\omega_\ell}\left(1+\mathcal{O}(\lambda^{-1})\right )
\end{equation}
where, $\Delta_\ell :=\gamma \zeta_\ell - \beta \phi_\ell(a_\ell)+\Pi_\ell$.

We take a moment to go through the various quantities appearing in this formula as we have introduced a few important subtleties regarding the different kinds of geometrical critical points that we neglected in the derivation above. The $\Pi_\ell$ contribution accounts for an extra phase in the Lorentzian intertwiners, see \cite{bianchi_lorentzian_2012,Barrett_2009}.
 The power $N$ is in general a half integer that depends on the rank of the hessian at the critical point and the combinatorics of the two-complex $\mathcal{C}$. The function $\mu(a)$ includes the summation measure over the spins and the Hessian evaluated at the critical point. The important point here is that neither the summation measure nor the Hessian scale with $\lambda$.

The estimation \eqref{final} is valid for all three types of possible geometrical critical points. If $\omega_\ell$ and $\textbf n_{\ell(\no)}$ specify a Lorentzian geometry, then 
\begin{equation}
	\beta=\gamma
\end{equation}
and
\begin{equation}
	\Pi_\ell = 
		\begin{cases} 
      		0 & \text{thick wedge}\\
      		\pi & \text{thin wedge}
   \end{cases}
\end{equation}
If $\omega_\ell$ and $\textbf n_{\ell(\no)}$ specify a degenerate geometry, then the dihedral angles $\phi_\ell(\omega_\ell,\textbf n_{\ell(\no)})$ either vanish or are equal to $\pi$, according to whether we are in a thick or thick wedge. By abuse of notation, we express this simply by setting $\beta=0$ in this case and keeping $\Pi_\ell$ defined as above.

If $\omega_\ell$ and $\textbf n_{\ell(\no)}$ specify a Euclidean geometry, then we have
\begin{equation}
	\beta=1
\end{equation}
and
\begin{equation}
	\Pi_\ell = 0 
\end{equation}
The function $\phi_\ell(s_c(v) ; \delta_\ell, \textbf n_{\ell(\no)})$ denotes the Palatini deficit angle. 

\medskip

This completes the analysis. The above results is the technique underlying the calculation presented in \cite{ChristodoulouCharacteristicTimeScales2018} to give an estimation of the bounce time for the black to white hole transition from spinfoams, which was based on a 2-complex without bulk faces. We expect future work to extend these results to also treat 2-complexes that include bulk faces. 

\section*{Chronology note}
The work presented here was mainly done during the period 2016-2019. It provides a self consistent presentation of the technique on which the calculation of the estimate for the bounce time of the black to white transition appeared in \cite{ChristodoulouCharacteristicTimeScales2018} was based. Some of the details presented here can be found scattered in the PhD manuscripts of coauthors MC and FDA, although different notation and conventions may be used. 

In order to avoid confusing the narrative, we did not cite above throughout the main body of the paper several works that have appeared after \cite{ChristodoulouCharacteristicTimeScales2018} appeared. A list of such works for which the results presented here might be relevant is \cite{Gozzini:2021kbt, Dona:2022dxs, Dona:2022yyn, Dona:2023myv, 
Dona:2018nev,
Bianchi:2018mml, Vidotto:2018wvr, DAmbrosio:2018wgv, Rovelli:2018cbg, Rovelli:2018hbk, Rovelli:2018okm, Alesci:2018loi, Kiefer:2019csi, Martin-Dussaud:2019wqc, Schmitz:2019jct, BenAchour:2020bdt, Piechocki:2020bfo, BenAchour:2020gon, Ong:2020xwv, Kelly:2020lec, Kelly:2020uwj, Zhang:2020lwi, DAmbrosio:2020mut, Schmitz:2020vdr, Barrau:2021spy, Mele:2021hro, Munch:2021oqn, Ansel:2021pdk, Soltani:2021zmv, Rignon-Bret:2021jch, Addazi:2021xuf,Husain:2022gwp, Barcelo:2022gii, Kazemian:2022ihc, Phat:2022xxw}.

\section*{Acknowledgements}

We acknowledge support of the ID\# 61466 grant from the John Templeton Foundation, as part of the ``Quantum Information Structure of Spacetime (QISS)'' project (\hyperlink{http://www.qiss.fr}{qiss.fr}). We thank Carlo Rovelli, Simone Speziale,  Pietro Dona and Pierre Martin-Dussaud for enlightening discussions during the course of this work.

\bibliography{PlanckStarLifetime,otherRefs,geomTransSpinfoams, mariosNewRefs,NewRefs,numerics}

\appendix

\setcounter{equation}{0}
\renewcommand\theequation{A.\arabic{equation}}
\section{Review of $SU(2)$ and $\slc$ Representation Theory}\label{appendix:App_RepTheory}
Let $\mathcal{V}^j$ with $j\in \frac{1}{2}\mathbb{N}$ be the vector space of homogeneous polynomials of degree $2j$ in two complex variables $\textbf{z}=(z_0, z_1)^\transpose\in\mathbb{C}^2$. More precisely, there exist coefficients $(a_0, \dots, a_{2j})\in\mathbb{C}^{2j+1}$ such that 
\begin{align}\label{eq:PolynomialAnsatz}
	P(\textbf{z}) = \sum_{k=0}^{2j}a_{k}\, z_0^k\, z_1^{2j-k},
\end{align}
which has the obvious property $P(\lambda \textbf{z}) = \lambda^{2j} P(\textbf{z})$ $\forall \lambda\in\mathbb{C}\backslash\{0\}$. In order to obtain a representation of $SU(2)$ on the vector space $\mathcal{V}^j$ we define the action of $h\in SU(2)$ as
\begin{align}\label{eq:GroupAction}
	h\rhd P(\textbf{z}) = P(h^\transpose \textbf{z}) \quad  \forall P\in\mathcal{V}^j.
\end{align}
and it is easy to verify that the two defining properties of a representation, i.e. 
\begin{align}
	\id \rhd P = P\quad\text{and}\quad (g_1 g_2)\rhd P = g_1\rhd(g_2\rhd P)
\end{align}
are satisfied. The so defined representation is finite-dimensional with $\dim \mathcal{V}^j=2j+1$ and one shows without much effort that it is also irreducible. Since all finite-dimensional irreducible representations of $SU(2)$ are isomorphic to one another we can relate the representation over $\mathcal{V}^j$ to the more familiar representation in terms of vectors $\ket{j m}\in\mathcal{H}^j$ by defining a linear map $\mathcal{I}:\mathcal{V}^j\rightarrow \mathcal{H}^j$ with the properties 
\begin{align}
	\mathcal{I}(P_m^j) = \ket{j m}\quad \text{and}\quad\mathcal{I}(h\rhd P)= h\rhd \mathcal{I}(P).
\end{align} 

This map allows us to determine a basis $P_m^j$ of $\mathcal{V}^j$. All we need is to do is to compare the action of $h\in SU(2)$ on $P_m^j$ and $\ket{j m}$. We therefore consider both sides of the equation
\begin{align}\label{eq:CompareEquation}
	h\rhd P_m^j = \mathcal{I}^{-1}(h\rhd\ket{j m})
\end{align}
separately and compare them in the end. Using the ansatz \eqref{eq:PolynomialAnsatz} for $P_m^j$ and the group action \eqref{eq:GroupAction} we get after some lengthy algebra
\begin{align}
	h\rhd P_m^j = \sum_{\vert l\vert \leq j}\sum_{\vert q \vert \leq j} a_{j+q} \left[\frac{(j+q)!(j-q)!}{(j+l)!(j-l)!} \right]^{\frac{1}{2}}  D_{lq}^j(h)z_0^{j+l}z_1^{j-l}.
\end{align}
for the left hand side of \eqref{eq:CompareEquation}. The evaluation of the right hand side is straightforward and we obtain
\begin{align}
	\mathcal{I}^{-1}(h\rhd \ket{jm})=\sum_{\vert r \vert \leq j}\sum_{\vert s \vert \leq j} a_{j+s}\,D_{r m}^j(h)\,z_0^{j+s}\,z_1^{j-s}.
\end{align}
Comparing these expressions term by term, i.e. by setting $l=s$ we obtain the condition
\begin{align}
	a_{j+q} \left[\frac{(j+q)!(j-q)!}{(j+s)!(j-s)!} \right]^{\frac{1}{2}} D_{sq}^j(h) \overset{!}{=} a_{j+s}\,D_{r m}^j(h).
\end{align}
This equation can only be satisfied for $s=r$ and $q=m$ from which it follows that
\begin{align}
	a_{j+m} \sqrt{(j+m)!(j-m)!} = a_{j+s} \sqrt{(j+s)!(j-s)!}.
\end{align}
Since $m$ is a fixed label we deduce that $a_{j+s}$ has to be of the form
\begin{align}
	a_{j+s} = \frac{C \delta_{sm}}{\sqrt{(j+m)!(j-m)!}}\quad\Longrightarrow\quad P_m^j(\textbf{z}) = \sum_{\vert s \vert \leq j} a_{j+s} \,z_0^{j+s}\,z_1^{j-s} = C \frac{z_0^{j+m}\,z_1^{j-m}}{\sqrt{(j+m)!(j-m!)}}
\end{align}
for some constant $C\in\mathbb{C}\backslash\{0\}$. This constant can easily be fixed by requiring that the basis $P_m^j$ be orthonormal with respect to an appropriate inner product on $\mathcal{V}^j$. When defining such an inner product we need to keep in mind convergence issues arising from integrating complex polynomials over $\mathbb{C}^2$. This excludes the Lebesgue measure and suggests the use of the measure
\begin{align}
	\frac{\ed^{-\innerp{\textbf{z}}{\textbf{z}}}}{\pi^2}\dd^4\textbf{z} := \frac{\ed^{-\innerp{\textbf{z}}{\textbf{z}}}}{\pi^2}\dd\text{Re}\left(z_0\right)\,\dd\text{Im}\left(z_0\right)\,\dd\text{Re}\left(z_1\right)\,\dd\text{Im}\left(z_1\right),
\end{align}
where the exponential damping factor ensures convergence. Hence, we can devise a well-defined and $SU(2)$ invariant inner product $\langle\cdot ,\cdot \rangle:\mathcal{V}^j\times\mathcal{V}^j\rightarrow \mathbb{C}$ by
\begin{align}\label{eq:InnerProductOnC2}
	\langle f, g \rangle_{\mathcal{V}^j} := \int_{\mathbb{C}^2}\dd^4\textbf{z}\, \frac{\ed^{-\innerp{\textbf{z}}{\textbf{z}}}}{\pi^2}\,\overline{f(\textbf{z})}\,g(\textbf{z}).
\end{align}

A nice property of this inner product is that it factorizes into separate integrations over $z_0$ and $z_1$. After a change to polar coordinates, one is left with simple integrals over Gaussian moments. It is therefore straight forward to check
\begin{align}
	\langle P_m^j, P_n^j\rangle = \vert C\vert^2\, \delta_{mn}\quad\Longrightarrow\quad \vert C\vert^2 = 1.
\end{align}
We choose $C=1$ for simplicity. The inner product \eqref{eq:InnerProductOnC2} allows us to write the resolution of identity on $\mathcal{H}^j$ in terms of the basis polynomials $P_m^j$ and this in turn will allow us to express the Wigner matrices in terms of complex polynomials. As an intermediate step, we define the ket 
\begin{align}
	\ket{j\,\textbf{z}} := \sum_{\vert m\vert \leq j} \overline{P_m^j(\textbf{z})}\ket{j\,m}
\end{align}
which by inspection has the property 
\begin{align}
	\innerp{j\,\textbf{z}}{j\, m} = P_m^j(\textbf{z}).
\end{align}
We can then write the identity on $\mathcal{H}^j$ as
\begin{align}
	\int_{\mathbb{C}^2} \dd^4\textbf{z}\, \frac{\ed^{-\innerp{\textbf{z}}{\textbf{z}}}}{\pi^2}\,\ket{j\,\textbf{z}}\bra{j\,\textbf{z}} = \id_{\mathcal{H}^j},
\end{align}
as can be checked by direct computation. Using the above resolution of identity we find for the Wigner matrices
\begin{align}
	D_{mn}^j(h) &= \bra{j\,m} h \ket{j\,n} = \bra{j\,m} \int_{\mathbb{C}^2} \dd^4\textbf{z}\, \frac{\ed^{-\innerp{\textbf{z}}{\textbf{z}}}}{\pi^2}\,h \ket{j\,\textbf{z}}\innerp{j\,\textbf{z}}{j\,n}\notag\\
	&=\int_{\mathbb{C}^2} \dd^4\textbf{z}\, \frac{\ed^{-\innerp{\textbf{z}}{\textbf{z}}}}{\pi^2}h\rhd\innerp{j\, m}{j\,\textbf{z}}\innerp{j\,\textbf{z}}{j\,n}\notag\\
	&=\int_{\mathbb{C}^2} \dd^4\textbf{z}\, \frac{\ed^{-\innerp{\textbf{z}}{\textbf{z}}}}{\pi^2} \overline{P_m^j(\textbf{z})}\,P_n^j(h^\transpose\textbf{z}).
\end{align}
To get from the first to the second line we used the fact that $h$ is acting on $\overline{P_m^j(\textbf{z})}$ inside $\ket{j\, \textbf{z}}$. In the third line we used $h\rhd \overline{P_m^j(\textbf{z})} = \overline{P_m^j(\overline{h}\textbf{z})} = P_m^j(h\overline{\textbf{z}})$ to perform the change of variables $\tilde{\textbf{z}}=\overline{h}\textbf{z}$ which produces the $h^\transpose \textbf{z}$ argument of $P_n^j$ (after dropping the tilde). As we will see in a moment, it is possible to generalize this method to the $\slc$ case.\newline
We recall that the principal series of $\slc$ is labeled by two parameters $\rchi \equiv (k, p)\in\mathbb{R}\times \frac{1}{2}\mathbb{Z}$. Let $\mathcal{V}^\rchi$ be the (infinite-dimensional) vector space of homogeneous meromorphic functions in two complex variables $\textbf{z}=(z_0, z_1)^\transpose\in\mathbb{C}^2$, where homogeneity now means
\begin{align}
	\Phi(\lambda \textbf{z}) = \lambda^{i k +p-1}\,\overline{\lambda}^{\,i k -p-1}\, \Phi(\textbf{z})\quad\forall\lambda\in\mathbb{C}\backslash\{0\}\text{ and }\forall \Phi\in\mathcal{V}^\rchi.
\end{align}

By defining the action of $g\in\slc$ on $\Phi\in\mathcal{V}^\rchi$ as
\begin{align}
	g\rhd \Phi(\textbf{z}) = \Phi(g^\transpose\textbf{z})
\end{align}
we obtain an infinite-dimensional irreducible representation of $\slc$ on $\mathcal{V}^\rchi$. Moreover, the $\mathcal{V}^\rchi$ representation splits into irreducible representations $\mathcal{V}^j$ of the $SU(2)$ subgroup
\begin{align}
	\mathcal{V}^\rchi \simeq \bigoplus_{j=\vert p\vert}^\infty \mathcal{V}^j,
\end{align}
where $j$ increases in integer steps. This fact allows us to define an injection at the fixed value $p = j$
\begin{align}
	\mathcal{J}&:\mathcal{V}^j\rightarrow\mathcal{V}^{(k, j)}\notag\\
	P(\textbf{z})&\mapsto \Phi(\textbf{z}) = \innerp{\textbf{z}}{\textbf{z}}^{i k-j-1} P(\textbf{z}),
\end{align}
which has indeed the correct homogeneity properties.

\pagebreak
\vfill

\setcounter{equation}{0}
\renewcommand\theequation{B.\arabic{equation}}

\section{The approximation $\vartheta_3\approx 1$}\label{appendixB}

Omitting details not relevant here and focusing only on one link for notational simplicity, the amplitude we would like to compute is given by
\begin{equation}
	W(A, \zeta) \simeq \sum_{j=0}^{\infty} \ed^{-t(A-j)^2+i\gamma \zeta j} \int_{\Omega}\dd\mu(g)\,\dd\nu(z)\,\ed^{j F(g, z)} \quad ,\quad g\in \text{SL}(2, \mathbb{C}), z\in \mathbb{C}^2.
\end{equation}
Using the splitting $j=A+s$ into fixed background geometry and fluctuations we get
\begin{equation}
	W(A, \zeta) \simeq \ed^{i \gamma \zeta A} \sum_{s=-\infty}^{\infty}\ed^{-t s^2 + i \gamma \zeta s} \int_{\Omega}\dd \mu(g) \,\dd\nu(z)\, \ed^{(A+s) F(g, z)}.
\end{equation}
Note that $\ed^{i \gamma \zeta A}$ is a pure phase (also in the general case when several links are present) and can therefore be neglected in what follows. Moreover, we used the approximation that $s\in (-\infty, \infty)$. The usual spin foam asymptotic analysis tells us that 
\begin{equation}
	\int_{\Omega}\dd \mu(g) \,\dd\nu(z)\, \ed^{(A+s) F(g, z)} \sim \ed^{-i \gamma \phi(g, z) A} \, \ed^{-i \gamma \phi(g, z) s}.
\end{equation}
We neglected here the Hessian and some numerical factors. Also, the phase $\ed^{-i \gamma \phi(g, z) A}$ can be neglected in what follows. We are hence left with 
\begin{equation}\label{eq:sum}
	W(A, \zeta) \sim \sum_{s=-\infty}^{\infty}\ed^{-ts^2 + i\gamma(\zeta-\phi)s}.
\end{equation}
Now, the sum \eqref{eq:sum} can be written down in closed form in terms of known functions as
\begin{equation}\label{eq:correctedresult}
	\sum_{s=-\infty}^{\infty}\ed^{-ts^2 + i\gamma(\zeta-\phi)s}  = \sqrt{\frac{\pi}{t}}\ed^{-\frac{\gamma^2}{4t}(\zeta-\phi)^2}\vartheta_3\left(-\frac{i\pi \gamma(\zeta-\phi)}{2t}, \ed^{-\frac{\pi^2}{t}}\right),
\end{equation}
where 
\begin{equation}\label{eq:thetafunction}
	\vartheta_3\left(u, q\right) = 1 + 2\sum_{n=1}^{\infty}q^{n^2}\cos(2 n u)
\end{equation}
is one of Jacobi's Theta functions. Hence, we have for the amplitude
\begin{equation}\label{ampl}
    W(A, \zeta) \sim  \sqrt{\frac{\pi}{t}}\ed^{-\frac{\gamma^2}{4t}(\zeta-\phi)^2}\vartheta_3\left(-\frac{i\pi \gamma(\zeta-\phi)}{2t}, \ed^{-\frac{\pi^2}{t}}\right)
\end{equation}
If we approximate $\vartheta_3$ with $1$ we obtain
\begin{equation}\label{approx ampl}
    W(A, \zeta) \sim  \sqrt{\frac{\pi}{t}}\ed^{-\frac{\gamma^2}{4t}(\zeta-\phi)^2}
\end{equation}

As we will see \eqref{approx ampl} is good an approximation to \eqref{ampl} in our setting. First, note that the two expressions have a significant qualitative difference: the former is periodic in $\zeta$ while the latter is not. We can read off the periodicity directly from the left hand side of \eqref{eq:correctedresult}:
\begin{equation}
	\ed^{i \gamma (\zeta-\phi)s} \Rightarrow \text{The period is } \frac{2\pi}{\gamma}.
\end{equation}
Let's examine carefully this periodicity. Since $\ed^{-t s^2}$ is always positive, we see that the maxima of the sum are located at
\begin{equation}
	\zeta_k = \phi + \frac{2\pi k}{\gamma} \quad , \quad k\in \mathbb{N}_0, \,\,\,\phi\in [0, 2\pi).
\end{equation}
The value of the maxima is then given by
\begin{equation}\label{eq:peak}
	\sqrt{\frac{\pi}{t}} \ed^{\frac{-k^2\pi^2}{t}}\vartheta_3\left(-\frac{i k \pi^2}{t}, \ed^{-\frac{\pi^2}{t}} \right) \equiv \sqrt{\frac{\pi}{t}} \vartheta_3\left(0, \ed^{-\frac{\pi^2}{t}} \right).
\end{equation}
In general we will have $K=\floor*{2\gamma}$ full periods in the interval $\zeta\in [0, 4\pi)$ and $M=1+\floor*{\gamma\left(2-\frac{\phi}{2\pi}\right)}$ maxima.
Now, we can exploit the freedom in restricting the value of the parameter $\gamma$. Since $\zeta\in [0, 4\pi)$ we find from $\frac{2\pi}{\gamma}\geq 4\pi$ that for $\gamma\leq\frac{1}{2}$ the periodicity of $W(A, \zeta)$ is not at all a problem. There will be less than one period and exactly one maximum in the interval $[0, 4\pi)$.

For completeness, we also note that $\phi$ essentially just moves around the maxima along the $\zeta$-axis, while $t$ determines their height and the spread of the Gaussians, as can be seen from \eqref{eq:peak}. It is also easy to see that the imaginary part of \eqref{eq:correctedresult} is exactly zero and that the real part is larger or equal to zero for all values of $\zeta, \phi$ and $t$.

In summary, when $\gamma \leq \frac{1}{2}$ we can safely use \eqref{approx ampl} instead of the more complicated result \eqref{ampl}. This is consistent with the fixing of the value of $\gamma$ that comes from calculating Black Hole entropy using LQG \cite{Ashtekar:1997yu,Ashtekar:2004nd,Meissner:2004ju}.  \newline

 \bigskip
Below we report graphical comparisons of the two expressions for the amplitude to illustrate the above reasoning.

\medskip

{\bf Example 1}: The choice of parameters is $t=0.01, \gamma=\frac{1}{3}$ and $\phi=3$. This means:
\begin{itemize}
	\item Period: $\frac{2\pi}{\gamma} = 6\pi$
	\item Number of full periods: $\floor*{2\gamma} = 0$
	\item Number of maxima: $1+\floor*{\gamma\left(2-\frac{\phi}{2\pi}\right)}=1$
	\item Location of maximum: $\phi+\frac{2\pi k}{\gamma} = 3$
	\item Height of maximum: $\sqrt{\frac{\pi}{t}} \vartheta_3\left(0, \ed^{-\frac{\pi^2}{t}} \right)= 17.72$
\end{itemize}

\includegraphics[width=0.85\textwidth]{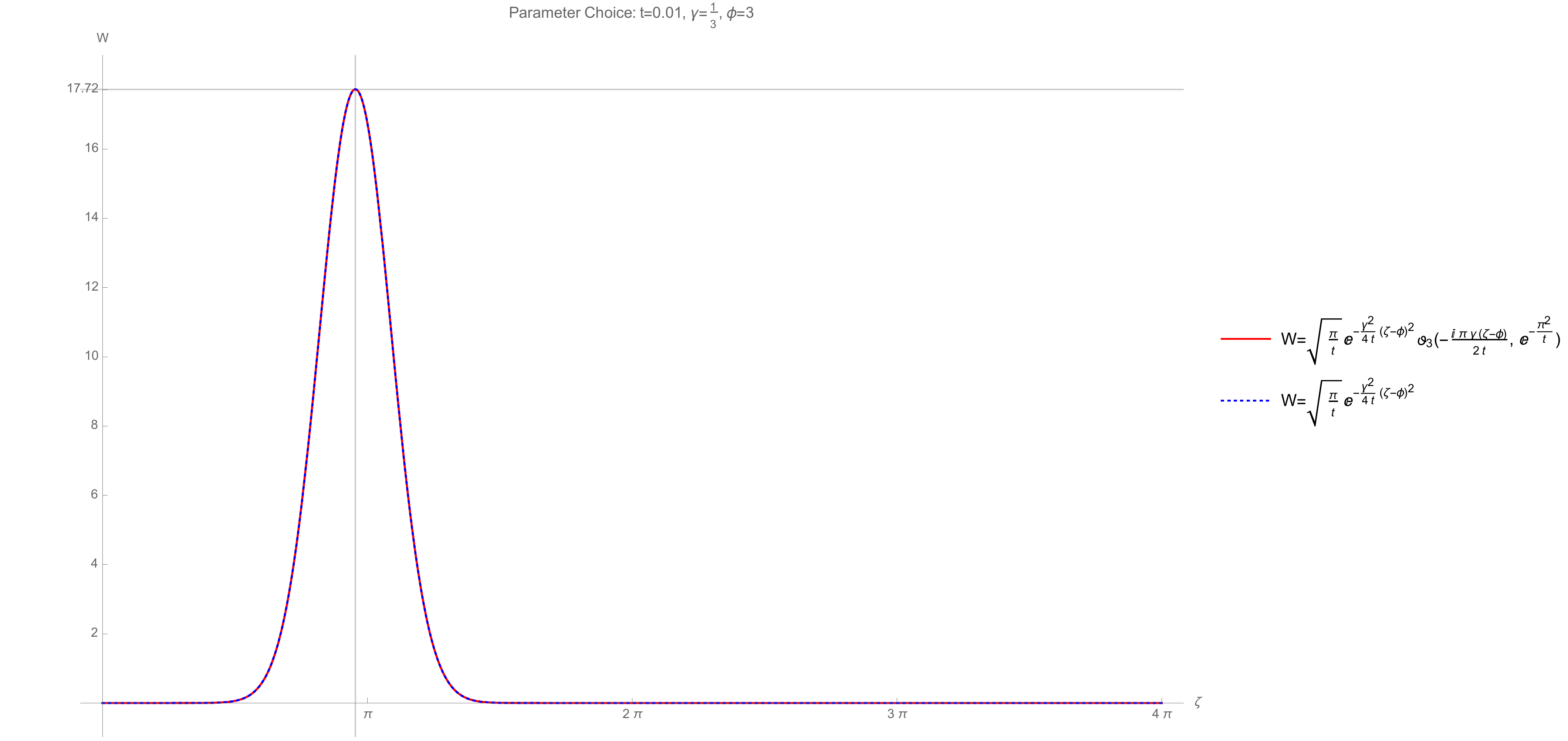}
\bigskip

{\bf Example 2}: The choice of parameters is $t=0.01, \gamma=2$ and $\phi = 1$.
\begin{itemize}
	\item Period: $\frac{2\pi}{\gamma}=\pi$
	\item Number of full periods: $\floor*{2\gamma} = 4$
	\item Number of maxima: $1+\floor*{\gamma\left(2-\frac{\phi}{2\pi}\right)}= 1+3 = 4$
	\item Location of maxima: $\{1, 1+\pi, 1+2\pi, 1+3\pi\}$
	\item Height of maxima: $\sqrt{\frac{\pi}{t}} \vartheta_3\left(0, \ed^{-\frac{\pi^2}{t}} \right)= 17.72$
\end{itemize}

\includegraphics[width=0.85\textwidth]{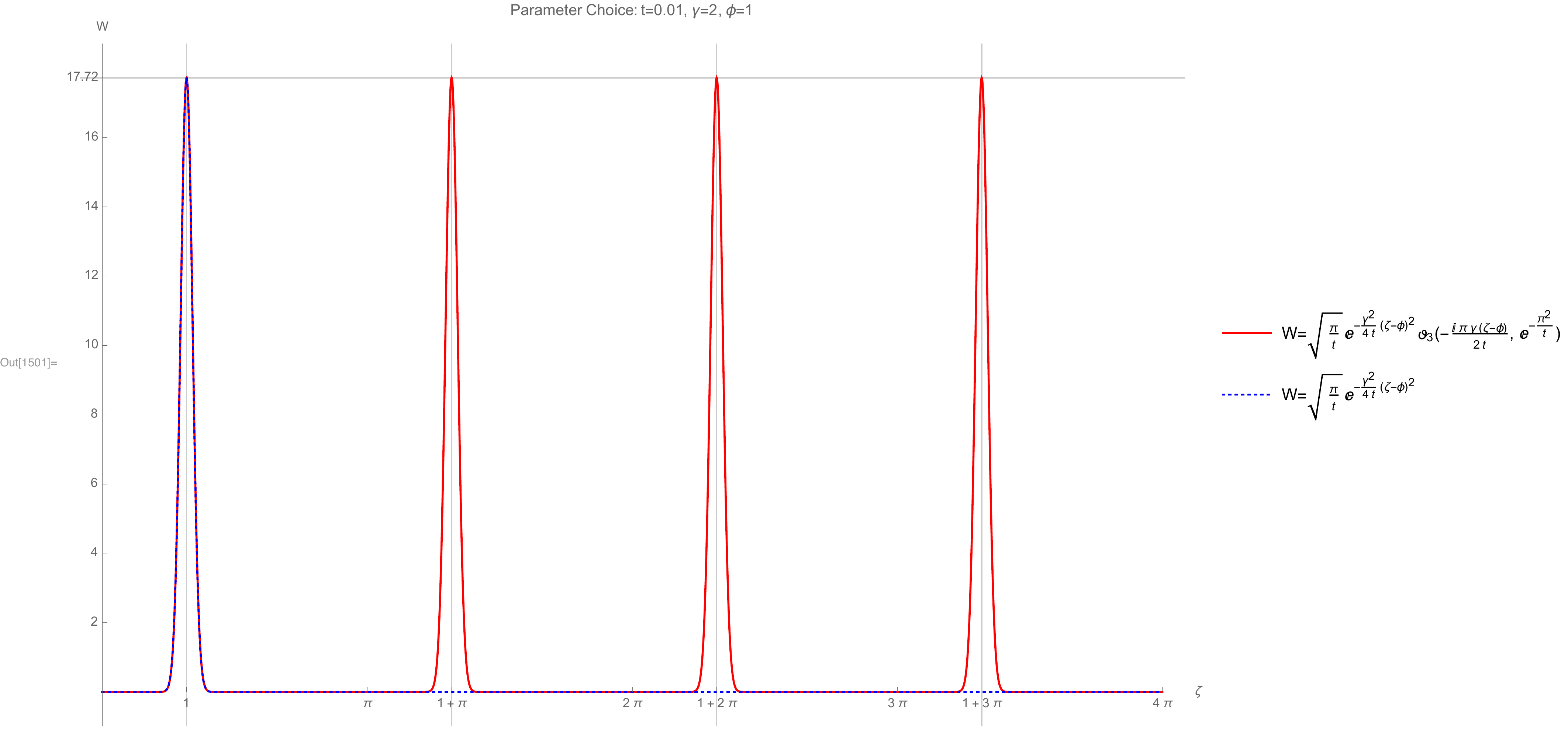}
\bigskip

{\bf Example 3}: The choice of parameters is $t=1, \gamma=\frac{3}{4}$ and $\phi=1$.
\begin{itemize}
	\item Period: $\frac{2\pi}{\gamma} = \frac{8\pi}{3}$
	\item Number of full periods: $\floor*{2\gamma} = 1$
	\item Number of maxima: $1+\floor*{\gamma\left(2-\frac{\phi}{2\pi}\right)}= 1+1=2$
	\item Location of maxima: $\{1, 1+\frac{8\pi}{3}\}$
	\item Height of maxima: $\sqrt{\frac{\pi}{t}} \vartheta_3\left(0, \ed^{-\frac{\pi^2}{t}} \right)=1.77$
\end{itemize}

\includegraphics[width=0.85\textwidth]{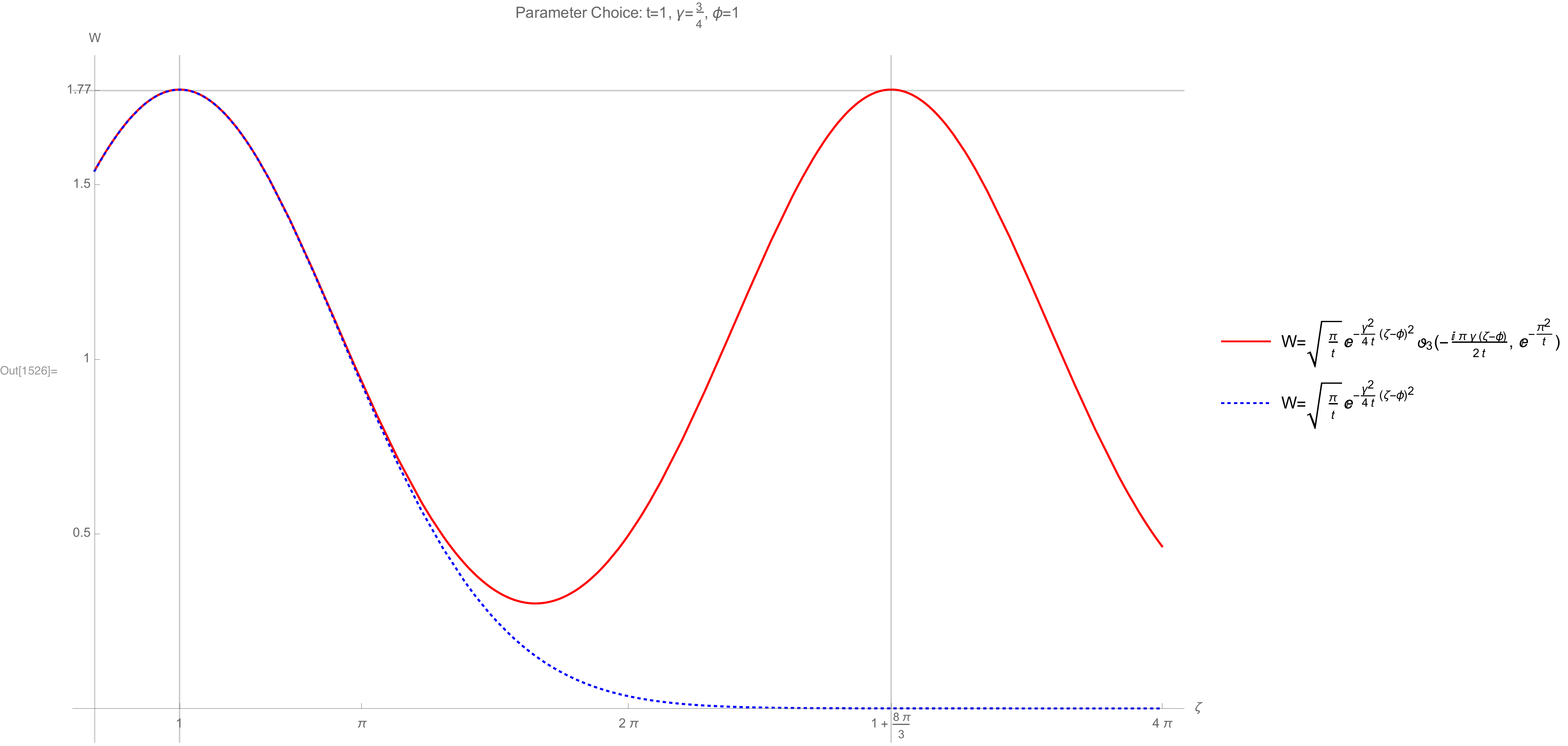}

\end{document}